\documentclass[modern]{aastex61}
\usepackage{amsmath}

\received{Received date}
\revised{Revised date}
\accepted{\today}
\submitjournal{ApJ}

\shorttitle{Radio observations of SN Master OT J120451.50+265946.6}
\shortauthors{Chandra et al.}

\begin{document}

\title{Type Ib supernova Master OT J120451.50+265946.6:  radio emitting shock with inhomogeneities  crossing through a dense shell }

\correspondingauthor{Poonam Chandra} 
\email{poonam@ncra.tifr.res.in}

\author[0000-0002-0844-6563]{Poonam Chandra}
\affiliation{National Centre for Radio Astrophysics, Tata Institute of Fundamental Research, PO Box 3, Pune, 411007, India}
\affiliation{Department of Astronomy, Stockholm University, AlbaNova, SE-106 91 Stockholm, Sweden}

\author{Nayana A. J.}
\affiliation{National Centre for Radio Astrophysics, Tata Institute of Fundamental Research, PO Box 3, Pune, 411007, India}

\author{C.-I. Bj{\"o}rnsson} 
\affiliation{Department of Astronomy, Stockholm University, AlbaNova, SE-106 91 Stockholm, Sweden}

\author{Francesco Taddia}
\affiliation{Department of Astronomy, Stockholm University, AlbaNova, SE-106 91 Stockholm, Sweden}

\author{Peter Lundqvist}
\affiliation{Department of Astronomy, Stockholm University, AlbaNova, SE-106 91 Stockholm, Sweden}

\author{Alak K. ray}
\affiliation{Homi Bhabha Centre for Science Education, 
Tata Institute of Fundamental Research, Mankhurd, Mumbai, 400088 India}

\author{Benjamin J. Shappee}
\affiliation{ Institute for Astronomy, University of Hawai'i, 2680 Woodlawn Drive, Honolulu, HI 96822, USA}

\begin{abstract}

We present radio observations of a Type Ib supernova (SN) Master OT J120451.50+265946.6. 
Our  low frequency Giant Metrewave Radio Telescope  (GMRT) data   taken when the SN was in  the optically thick phase for observed frequencies  reveal inhomogeneities in the structure of the radio emitting region.  
The high frequency  Karl G. Jansky Very Large Array data indicate that the shock is crossing through   a dense 
shell between  $\sim$ 47 to  $\sim 87$ days.   
 The data   $\ge 100$ days onwards are reasonably well fit with the inhomogeneous synchrotron-self absorption model.
Our model  predicts  that the inhomogeneities should smooth out  at late times. Low frequency GMRT observations at late epochs will test this prediction. Our findings suggest the importance of obtaining well-sampled  wide band radio data in order to understand the intricate nature of the radio emission from young supernovae.

\end{abstract}


\section{Introduction} \label{sec:intro}

Core-collapse supernovae (SNe) are energetic explosions that mark the death of massive stars with masses M $>$ 8M$_{\odot}$. Type Ib and Ic SNe  (SNe Ib/c)
are characterised by  absence of H and/or   He lines  in their optical spectra,  suggesting  that the outer H/He envelopes of the progenitor star were ejected before the 
explosion in these SNe \citep{woosley2002}. 
These are collectively called  stripped 
envelope SNe \citep{clocchiatti1997}.   Among all local SNe (distance $d< 60$\,Mpc), 19\% belong to the class of SNe Ib/c \citep{li2011}.

 The proposed plausible progenitors of  SNe Ib/c are Wolf-Rayet (W-R) stars that strip most of their outer envelope/s via strong 
stellar winds, and  stars in close binary systems where H/He envelopes of the progenitor star are stripped via binary interactions \citep{ensman1988}.  
The physical mechanisms
by which the progenitors of SNe Ib/c lose their H/He envelopes before the explosion and the time scales involved in this process remain open questions. 

 The progenitor systems of SNe Ib/c are poorly constrained from direct detection efforts from pre-explosion images \citep{smartt2009}. However, independent constraints on the properties of SN progenitors can be obtained by studying the interaction of SNe with their circumstellar medium (CSM), formed due to mass lost from the
 progenitor systems during their evolutionary phases. The  mass-loss during the evolution can be constant creating steady stellar winds,  or can occur via episodic events \citep{dopita1984} creating a complex density field around the star.  Radio emission, emitted via the synchrotron mechanism,  is one of  the best signatures to study  the SN ejecta interaction with the CSM, and probes the properties of the CSM \citep{chevalier1982}.

In this work, we report the radio observations of SN  Master OT J120451.50+265946.6 (hereafter SN\,J1204).
SN\,J1204 was discovered on on 2014 October 28.87 (UT) by \citet{gress2014}  with  an optical magnitude  $m_v=13.9$ and at a position $\alpha_{\rm J2000}$ = 12$^{\rm h}$04$^{\rm m}$51.5$^{\rm s}$, $\delta_{\rm J2000}$ = +26$^{\circ}$59$^{\prime}$46.6$^{\prime \prime}$.  The SN exploded in a galaxy  NGC\,4080 at a distance  $d \sim$ 15 Mpc \citep{karachentsev2013}.
 SN\,J1204 was classified as a type Ib SN based on the spectrum obtained with the  2m Himalayan 
 Chandra Telescope (HCT) of the Indian Astronomical Observatory  on 2014 Oct 29.0 UT \citep{srivastav2014}. Later \citet{terreran2014} confirmed the classification using the spectrum obtained with the Asiago 1.82 m Telescope. 
The expansion velocity was found to be 8300 km s$^{-1}$ from HeII absorption feature in the HCT spectrum 
 \citep{srivastav2014}. SN\,J1204 observations with the X-ray telescope (XRT)  on-board \textit{Swift}  during  2014 October 29--November 04 yielded upper limits
 \citet{margutti2014}.

 The radio emission from SN\,J1204 was first detected at 5\,GHz with   the Karl G. Jansky  Very Large Array (VLA) on 2014 October 31.5 with a flux density of 3.00 $\pm$ 0.02 mJy \citep{kamble2014}.  The Giant Metrewave Radio Telescope (GMRT) detected the radio emission  at the SN position
 at 1390 MHz with a flux density of 1.56 mJy on 2014 November 26.09 UT \citep{poonam2014}. 
In this paper, we present the results of an extensive radio follow up of SN\,J1204 with the GMRT, combined with the publicly available  data with the VLA.
Thus covering  a frequency range of 0.33 to 24 GHz for more than 1000 days  since discovery,
 we carry out detailed spectral and temporal modeling of the SN and derive the nature of the radio emission.

The paper is organised as follows: In \S \ref{sec:obs}, we summarise the observations at various wavebands and procedures for data analysis. 
We use the optical photometry data to best constrain the epoch of explosion in \S \ref{sec:explosion}. In \S \ref{sec:model}, we attempt to fit the data with the standard models 
of radio emission and show that
this model  is incapable of fitting the data. We develop and fit an inhomogeneous model to the radio data in \S \ref{sec:non-standard}. In addition, in this section, we  also discuss  the presence of a dense shell.  Finally we discuss our results, compare the properties of SN\,J1204 with other stripped-envelope  SNe and summarise our main conclusions in \S \ref{discussion}.

\section{Observations and Data analysis} \label{sec:obs}

\subsection{Radio Observations}
 SN\,J1204 has been extensively observed in the radio bands using the GMRT and the VLA. Below we describe the observations and procedures for data analysis.

\subsubsection{GMRT Observations}\label{sec:gmrt}

\startlongtable
\begin{deluxetable*}{ccccccc}
\tablecaption{Details of GMRT observations of SN Master OT J120451.50+265946.6 (SN\,J1204)  \label{tab:gmrt}}
\tablehead{
\colhead{Date of} & \colhead{Age\tablenotemark{a}} & \colhead{Frequency} & \multicolumn2c{Flux density} & \colhead{Map rms} & \colhead{Resolution} \\
\cline{4-5}
\colhead{observation} & \colhead{} & \colhead{} &  \colhead{SN} &  \colhead{Test source\tablenotemark{b}} & \colhead{} & \colhead{} \\
\colhead{(UT)} & \colhead{(day)} & \colhead{(GHz)} & \colhead{(mJy)} &  \colhead{(mJy)} & \colhead{($\mu$Jy\,beam$^{-1}$)} & \colhead{($''\times''$)}
}
\startdata
2014 Nov 26.1 & 61.1 & 1.39 &  1.54$\pm$0.17 & 6.99 $\pm$ 0.70  & 41 & 2.30 $\times$ 1.95 \\
2014 Dec 14.0 & 79.0 & 1.39 &  1.63$\pm$0.18 & 6.40 $\pm$ 0.65 & 44 & 2.28 $\times$ 1.96 \\
2015 Jan 29.8 & 125.8 & 1.39 &  3.18$\pm$0.33 & 5.85 $\pm$ 0.59 & 50 & 3.88 $\times$ 2.03 \\
2015 May 30.4 & 246.4  & 1.39 &  1.65$\pm$0.19 & 6.09 $\pm$ 0.62 & 62 & 3.65 $\times$ 2.73 \\
2015 Jul 04.4 & 281.4  & 1.39 &  1.33$\pm$0.17 & 6.37 $\pm$ 0.64 & 46 & 2.24 $\times$ 1.84 \\
2015 Jul 17.5 & 294.5  & 1.39 &  1.14$\pm$0.13 & 6.28 $\pm$ 0.63 & 40 & 2.20 $\times$ 2.01 \\
2015 Aug 17.5 & 325.5  & 1.39 &  0.90$\pm$0.11 & 5.85 $\pm$ 0.59 & 44 & 2.59 $\times$ 2.06 \\
2015 Aug 21.6 & 326.6  & 1.39 &  1.01$\pm$0.15 & 5.64 $\pm$ 0.58 & 60 & 3.88 $\times$ 2.02 \\
2015 Sep 18.5 & 357.5  & 1.39 &  0.85$\pm$0.17 & 6.96 $\pm$ 0.72 & 80 & 8.58 $\times$ 3.06 \\
2017 Apr 21.6 & 938.6  & 1.39 &  0.22 $\pm$ 0.06   & 5.31 $\pm$ 0.54 & 40 & 3.88 $\times$ 2.02 \\
\hline
2014 Dec 03.0 & 68.0 & 0.61 & 0.47 $\pm$ 0.12 & 13.14 $\pm$ 1.32 & 68 & 5.14 $\times$ 4.18 \\
2014 Dec 21.9 & 86.9 & 0.61& 0.53 $\pm$ 0.11& 15.91 $\pm$ 1.60 & 60 & 6.60 $\times$ 4.30 \\
2015 Feb 03.1 & 130.1 & 0.61 & 1.00$\pm$0.18 & 15.78 $\pm$ 1.58 & 59 & 9.03 $\times$ 4.74 \\
2015 Mar 14.8 & 169.8  & 0.61   &  1.56$\pm$0.19 & 15.29 $\pm$ 1.53 & 55 & 5.00 $\times$ 4.22 \\
2015 May 01.6 & 217.6  & 0.61   &  1.98$\pm$0.25 & 16.37 $\pm$ 1.65 & 78 & 5.76 $\times$ 4.30 \\
2015 Jun 06.4 & 253.4  & 0.61   &  2.48$\pm$0.28 & 17.38 $\pm$ 1.74 & 76 & 6.34 $\times$ 4.60 \\
2015 Jul 10.7 & 287.7  & 0.61   &  2.21$\pm$0.37 & 13.86 $\pm$ 1.40 & 98 & 4.48 $\times$ 3.76 \\
2015 Jul 26.8 & 303.8  & 0.61   &  2.26$\pm$0.27 & 15.75 $\pm$ 1.58 & 84 & 5.50 $\times$ 4.45 \\
2015 Sep 25.3 & 364.3  & 0.61   &  1.92$\pm$0.23 & 15.94 $\pm$ 1.60 & 67 & 6.20 $\times$ 4.70 \\
2017 Apr 29.6 & 946.6  & 0.61   &  0.57 $\pm$ 0.11  & 16.41 $\pm$ 1.64 & 64 & 6.91 $\times$ 4.43 \\
\hline
2015 Jul 13.5 & 290.5 & 0.33 & $\le 1.086$ & $23.13\pm3.49$ & 362 & 9.33 $\times$ 8.91 \\
2015 Sep 21.3 & 360.3 & 0.33 & $\le3.87$  & $22.89\pm3.69$ & 1290 & 11.17 $\times$ 7.65 \\
2016 Oct 21.0 & 756.0 & 0.33 & $1.85\pm0.45$ & $25.10\pm3.79$ & 245 & 16.42 $\times$ 8.47 \\
2017 Apr 27.5 & 944.5 & 0.33 & $0.72\pm0.31$ & $30.06\pm4.53$ & 263 & 12.03 $\times$ 8.17 \\
2017 Nov 27.0 & 1158.0 & 0.33 & $0.89\pm0.38$ & $29.65\pm4.47$ & 248 & 11.58 $\times$ 8.00 \\
\enddata
\tablenotetext{a}{The age is calculated assuming 2014 September 26 (UT) as the date of explosion (see \S \ref{sec:explosion}).}
\tablenotetext{b}{Nearby constant flux density test source at a position of  $\alpha_{\rm J2000}$ = 12$^{\rm h}$04$^{\rm m}$29.01$^{\rm s}$, $\delta_{\rm J2000}$ = +27$^{\circ}$03$^{\prime}$45.3$^{\prime \prime}$ (see \S \ref{sec:gmrt}).}
\tablecomments{The tabulated uncertainties in all flux density  measurements are obtained using AIPS task JMFIT plus 10\% uncertainties added in quadrature for 1390 and 610 MHz bands, and
15\% for 325 MHz band.}	
\end{deluxetable*}

\begin{deluxetable*}{lccccc}[h!]
	\centering
	\tablecaption{Details of VLA observations of Master OT J120451.50+265946.6  \label{tab:vla}}
	\tablecolumns{6}
	\tablewidth{0pt}
	\tablehead{
		\colhead{Date of } & \colhead{Age\tablenotemark{a}} & \colhead{Frequency} & \colhead{VLA Array} & \colhead{Flux density} & rms \\
		\colhead{Observation (UT)} & \colhead{(day)} &
		\colhead{(GHz)} & \colhead{Configuration} & \colhead{mJy} & \colhead{$\mu$Jy\,beam$^{-1}$}
	}
	\startdata
	2014 Oct 31.5  & 35.5  & 4.799 & C  & 2.938$\pm$0.151 & 10    \\
	- & -  & 7.099 &  C & 2.213$\pm$0.117  & 9  \\	
	2014 Nov 12.5 & 47.5   & 1.515 & C & 1.420$\pm$0.123 & 80  \\ 
	-    & - & 4.799 & C & 2.520$\pm$0.131 & 13   \\   
	-  & - & 7.099 & C & 1.858$\pm$0.097 & 12   \\    
	-  & - & 13.499 & C & 0.886$\pm$0.046  & 14   \\   
	- & - & 15.999 & C & 0.696$\pm$0.040  & 18   \\
	- & - & 19.199  & C &0.583$\pm$0.038 & 22   \\
	- & - & 24.499 & C & 0.429$\pm$0.039 & 27   \\    
	2015 Jan 08.4  & 103.4   & 1.515 & CnB & 2.404$\pm$0.142 & 32    \\
	-  & - & 2.529 & CnB & 3.238$\pm$0.165 & 30     \\
	-  & - & 3.469 & CnB & 2.707$\pm$0.137 & 15     \\
	-  & - & 4.799 & CnB & 2.151$\pm$0.109 & 19   \\
	-  & - & 7.099 & CnB & 1.495$\pm$0.076 & 22    \\
	-  & - & 13.499 & CnB & 0.781$\pm$0.041 & 22   \\
	-  & - & 15.999 & CnB & 0.622$\pm$0.032 & 12    \\
	- & - & 19.199 & CnB & 0.495$\pm$0.036  & 16   \\
	- & - & 24.499 & CnB & 0.367$\pm$0.030  & 18 \\		
	2015 Apr 22.1  & 207.1 & 1.515 & B & 1.793$\pm$0.111 & 14    \\
	-  & - & 2.529 & B & 1.186$\pm$0.064 & 14    \\
	-  & - & 3.469  & B &0.886$\pm$0.051 & 13   \\
	-  & - & 4.799 & B & 0.683$\pm$0.038 & 14    \\
	-  & - & 7.099 & B & 0.476$\pm$0.028 & 13   \\
	-  & - & 13.499 & B & 0.224$\pm$0.021 & 13 \\
	-  & - & 15.999 & B & 0.162$\pm$0.024 & 13   \\
	-  & - & 19.199 & B & 0.181$\pm$0.018 & 17    \\
	-  & - & 24.499 & B & 0.124$\pm$0.019 & 21   \\	
	2015 July 31.1  & 307.1  & 1.515 & A & 1.010$\pm$0.130 & 17      \\
	-  & - & 2.529 & A & 0.500$\pm$0.036  &16     \\
	-  & - & 3.469 & A & 0.466$\pm$0.038  & 16   \\
	-  & - & 4.799 & A & 0.452$\pm$0.032  & 13    \\
	-  & - & 7.099 & A & 0.241$\pm$0.019  &12    \\
	-  & - & 8.599 & A & 0.193$\pm$0.017  &13   \\
	-  & - & 10.999 & A &  0.181$\pm$0.014 &16    \\
	\enddata
	\tablenotetext{a}{The age is calculated using 2014 September 26 (UT) as the date of explosion.}
	\tablecomments{Here the uncertainties in the flux density measurements  reflect the map uncertainty and 5\% uncertainty in flux density estimation added in quadrature.}
\end{deluxetable*}

We started the GMRT observations of SN\,J1204 starting 2014 November 26.09 UT and  continued monitoring observations for about 3 years. The observations covered the frequency bands of 
 1390, 610 and 325 MHz.
 The observations were carried out in  total intensity mode   (Stokes I),  and the data were acquired with an integration time of 16.1 sec. The observed bandwidth  was 33 MHz,  split into 256 channels at all
observed  frequencies and epochs. A flux calibrator (either 3C286 or 3C147) was observed once during  each observing run to calibrate the amplitude gains of individual antennas. Phase calibrators 
 (J1125+261, J1227+365 or J1156+314) were observed every $\sim35$ minutes for  $\sim5$ minutes throughout each observing run to correct for phase variations due to atmospheric fluctuations. The data were analysed using  the Astronomical Image Processing System (AIPS). Initial flagging and calibration were done using the software FLAGCAL, developed for automatic flagging and calibration for the GMRT data \citep{prasad2012}. The flagged and calibrated data were closely inspected, and further flagging and calibration were done manually until the data quality looked satisfactory. The calibration solutions obtained for  a single channel were applied to the full bandwidth. Instead of averaging the full bandwidth,  only a few channels (20 channels for 1390 MHz band, 15 channels for 610 MHz band and 7 channels for 325 MHz band) were averaged together to avoid bandwidth smearing.  Fully calibrated data of the target source was imaged using the AIPS task IMAGR. A few rounds of phase self calibration and  one round
 of amplitude plus phase (a\&p) self calibration were performed.   During the a\&p  self-calibration, the voltage gains were normalized to unity to minimize the drifting of the flux-density scale. 
  In addition, the   flux densities of SN J1204 and the test source were measured between the a\&p and the last phase self-calibration rounds and they were found to be consistent.  The flux density of the SN was measured
  by fitting a Gaussian at the SN position using the AIPS task JMFIT.   This is a standard procedure followed for the GMRT data analysis  \citep[e.g. ][and references therein]{ck17}. 
 The data on 2015 April 25 were found to be corrupted and we did not use it for further analysis. The details of GMRT observations and the SN flux densities are summarised in Table \ref{tab:gmrt}.  

\begin{figure}
 \centering
 \includegraphics*[width=0.45\textwidth]{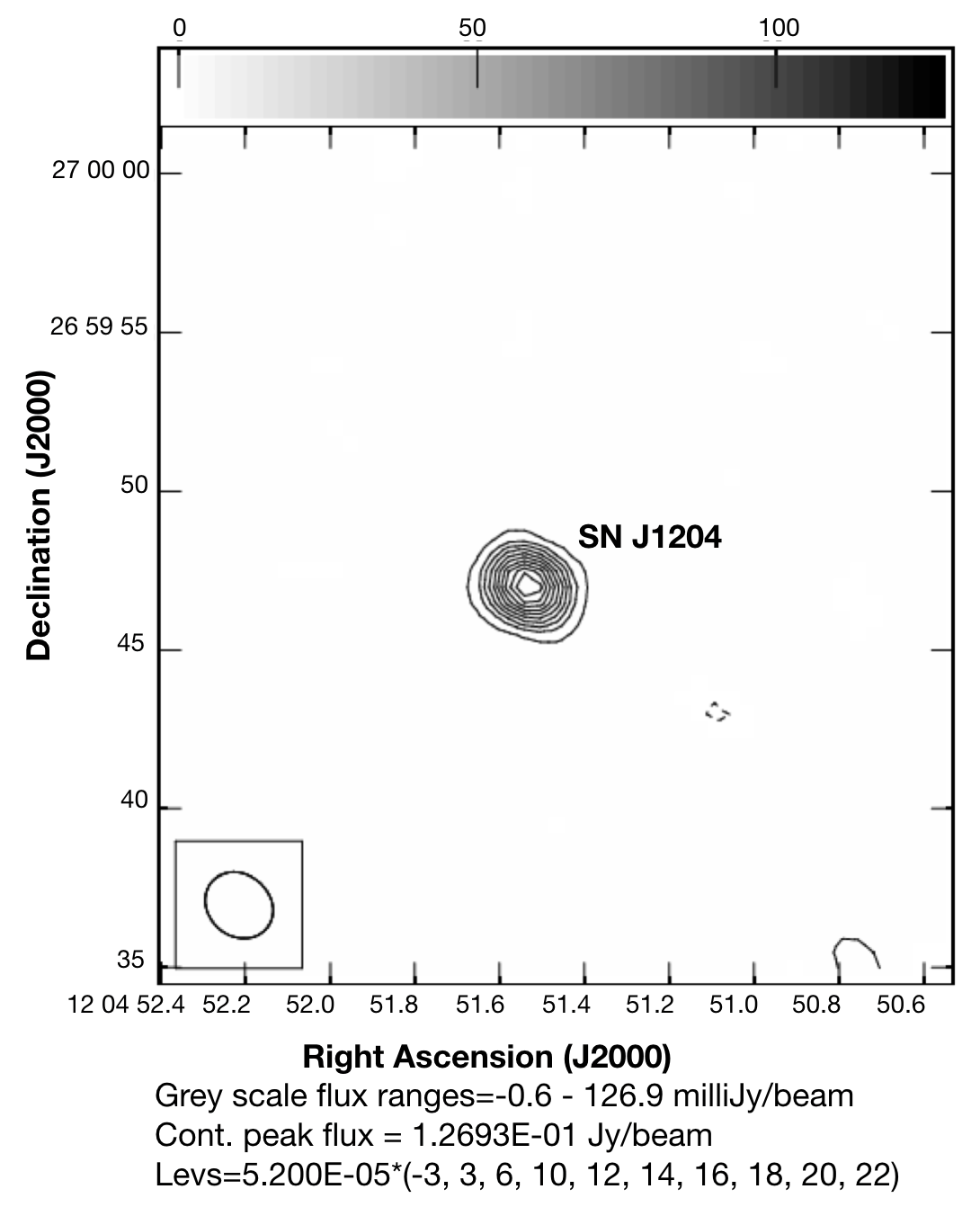}
 \caption{   Contour plot map of the SN\,J1204 field at the GMRT 1387 MHz band. The contours are marked as 3, 6, 10, 12, 14, 16, 18, 20, 22 times the map uncertainty very near to the SN position, respectively. One can note that there is no 
 contaminating source at the SN position  within $3\sigma$.}
     \label{contour}
 \end{figure}

    To estimate whether there is a contribution from the host galaxy at the SN position, in  Fig. \ref{contour}  we plot the the contour map for the SN field. We do not find any contamination 
    by the host galaxy at the   SN position within 3$\sigma$ noise.

 To confirm if the variability seen in the SN flux density at various epochs is real, we chose a nearby non-variable test source  NVSS J120428+270343 at a position of  $\alpha_{\rm J2000}$ = 12$^{\rm h}$04$^{\rm m}$29.01$^{\rm s}$, $\delta_{\rm J2000}$ = +27$^{\circ}$03$^{\prime}$45.3$^{\prime \prime}$. 
The source was selected such that its flux density was  found to be constant within the uncertainties of 10\% of the source flux between the NVSS 
 \citep{nvss} and
 the FIRST \citep{first1, first2} surveys.
The flux density of the test source in our observations is constant within $\sim10$\% errors at 1390 and 610 MHz bands and within $\sim15$\% at 325 MHz band at various epochs (Table \ref{tab:gmrt}).  This is roughly consistent with the uncertainties in the
 GMRT data due to systematic errors \citep{ck17}. Thus for fitting the data, we add 10\% uncertainty in quadrature  at 1390 and 610 MHz bands and 
 15\% at 325 MHz bands.

\subsubsection{VLA Observations}

We  analysed the publicly available archival VLA data for SN\,J1204 at five epochs from 2014 October 31 to 2015 July 31  spanning a frequency range of 1--24\,GHz.   
The data had bandwidth of $\approx$ 1 GHz split into 8 spectral windows at 1.515\,GHz,  and  $\approx$ 2 GHz  split into 16 spectral windows for all other frequencies. 
We carried out the VLA data analysis using standard packages within the Common Astronomy Software Applications package \citep[CASA,][]{casa}. 
CASA task `clean' was used to make images of the data. The details of the VLA observations, array configurations and the flux densities of the SN are summarised in Table \ref{tab:vla}.  
We add 5\% error in the quadrature for the VLA data, a typical uncertainty  in the flux density calibration scale at the  observed 
frequencies \footnote{\url{https://science.nrao.edu/facilities/vla/docs/manuals/oss/performance/fdscale}}.

 \begin{figure}
\centering
\includegraphics*[width=0.69\textwidth]{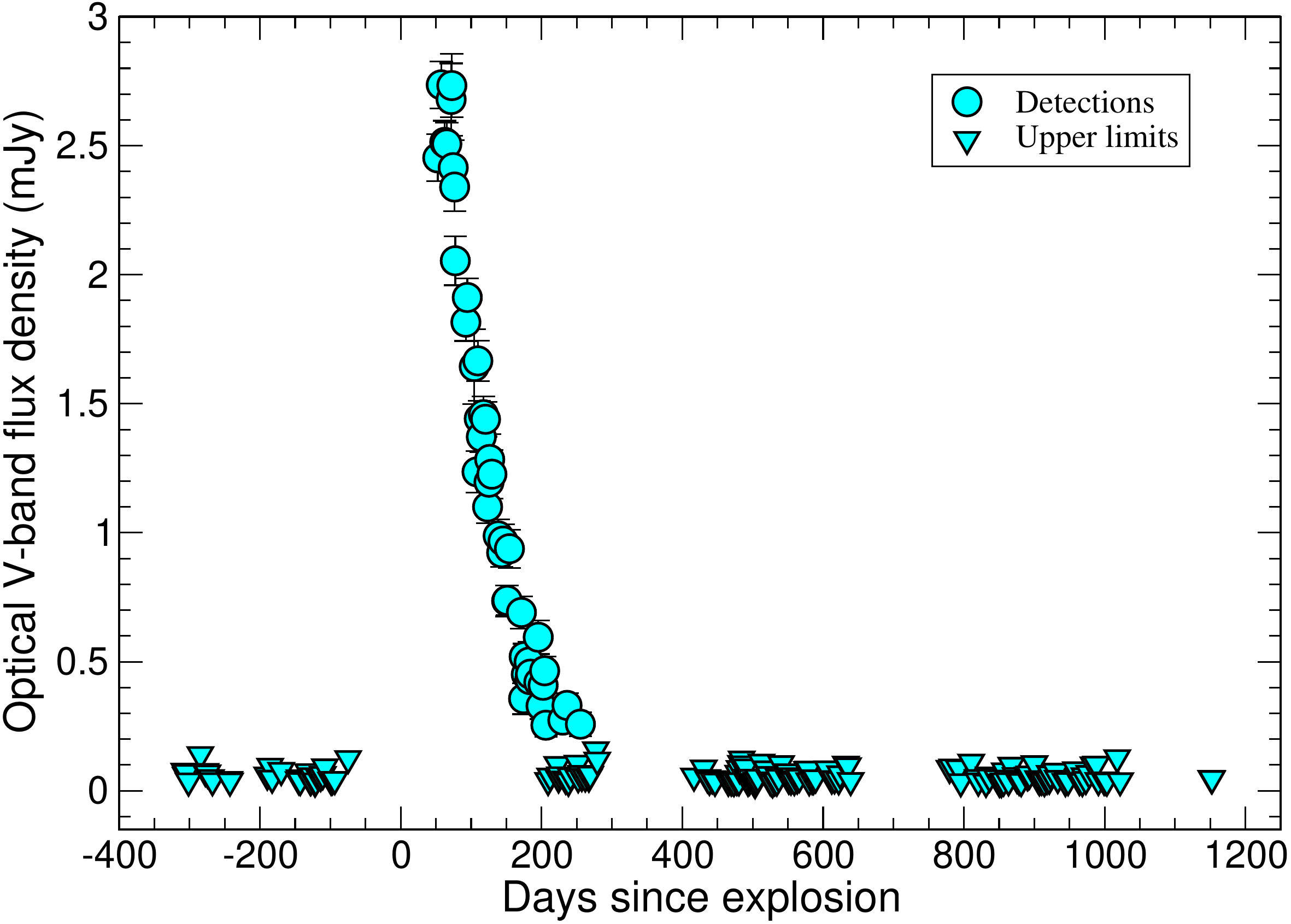}
\caption{ ASAS-SN optical data for SN\,J1204  covering epochs 400 days before the SN explosion till 1200 days post  explosion. The detections are indicated by circles 
and the  3-$\sigma$ upper limits are indicated by triangles.  Zero time corresponds to
the SN explosion date, i.e. 2014 Sep 26 UT.   }
    \label{optical1}
\end{figure}

\subsection{Optical Observations}

We used the data from All-Sky Automated Survey for Supernovae \citep[ASAS-SN,][]{shapee14} in V-magnitude band,  covering pre-explosion phase up to 400 days before the
assumed explosion date until 1200 days 
post-explosion.  ASAS-SN images were processed by a fully automatic ASAS-SN pipeline using the  Interactive Spectral Interpretation System \citep[ISIS; ][]{isis} image subtraction package \citep{alardlupton1998,alard2000}.  For this the stacking was done on the three dithered images before  carrying out the photometry. For the subtraction, the stacked images were
subtracted from a good reference image.    We performed aperture photometry on the subtracted images using  Image Reduction and Analysis Facility \citep[IRAF; ][]{iraf1,iraf2} `apphot' package and calibrated the results using the AAVSO photometric All-sky Survey  \citep[APASS;][]{henden2015}. Some of the data points were found to be affected by clouds in the FOV or by the 
incidence of cosmic rays;  and these data were discarded.  The ASAS-SN detections and 3$\sigma$ limits are shown in Fig. \ref{optical1}.  SN\,J1204 is, unfortunately, only detected after a seasonal gap which resulted in missing the peak of the SN light curve.

\begin{deluxetable*}{lccccccc}
\tablecaption{X-ray observations of Master OT J120451.50+265946.6  (SN\,J1204) \label{xray}}
\tablehead{
\colhead{Date of} & \colhead{Age\tablenotemark{a}} & \colhead{Telescope} &  \colhead{ObsID} &  \colhead{exposure} &  \colhead{Count rate\tablenotemark{b}} 
& \colhead{Flux\tablenotemark{b}}  & \colhead{Luminosity\tablenotemark{b}} \\
\colhead{obsn (UT)} & \colhead{day} & \colhead{} &  \colhead{} &  \colhead{(ks)} & \colhead{(counts s$^{-1}$)} & \colhead{(erg s$^{-1}$ cm$^{-2}$)} & 
\colhead{(erg $^{-1}$)}
}
\startdata
2014 Oct 29.50  & 33.50 & {\it Swift}-XRT & 00033511001 & 1.98 & $<7.66\times10^{-3}$ & $<2.62\times10^{-13}$ & $<7.06\times10^{39}$\\
2014 Nov 02.83   & 36.83 &  {\it Swift}-XRT &  00033511002 & 1.61 & $<7.05\times10^{-3}$ & $<2.41\times10^{-13}$  & $<6.49\times10^{39}$ \\
2014 Nov 04.30    & 38.30 &  {\it Swift}-XRT & 00033511003	 & 1.61 & $<7.05\times10^{-3}$ & $<2.41\times10^{-13}$  & $<6.49\times10^{39}$ \\
2014 Nov 16.88  & 50.88 & {\it Chandra}-ACIS &  16006 & 9.65 &   $<2.36\times10^{-4}$ & $<8.06\times10^{-15}$  & $<2.17\times10^{38}$\\
2015 Nov 25.74  & 424.74  & {\it Swift}-XRT & 00084388001 & 4.59 & $<2.46\times10^{-3}$ & $<8.41\times10^{-14}$  & $<2.27\times10^{39}$\\
2017 Jan 21.83 & 847.83 & {\it Swift}-XRT & 00033511001 & 1.65 & $<7.01\times10^{-3}$ & $<2.40\times10^{-13}$   & $<6.46\times10^{39}$\\
2017 Oct 29.69  & 1128.69 & {\it Swift}-XRT & 00084388002 & 0.55 & $<2.11\times10^{-2}$ & $<7.21\times10^{-13}$  & $<1.94\times10^{40}$\\
\enddata
\tablenotetext{a}{The age is calculated using 2014 September 26 (UT) as the date of explosion.}
\tablenotetext{b}{The values are in the  energy range 0.3--10\,keV.}
\end{deluxetable*}

\subsection{X-ray Observations} 

SN\,J1204 was observed with the  {\it Swift}-XRT covering a period  almost up to 1000 days  since its discovery.  In addition,
{\it Chandra} observed it for around 10\,ks on 2014 Nov 16.
We analysed the publicly available  archival data from  both the telescopes (see Table \ref{xray}). 

For the {\it Chandra} data analysis, the 
Chandra Interactive Analysis of Observations software \citep[CIAO;][]{frus06}
was used.
We extracted spectra, response and ancillary matrices using 
the CIAO task
{\it specextractor}. The 
CIAO version 4.9 along with CALDB version 4.7.6 was used for this
purpose. 

The {\it Swift}-XRT spectra and response matrices were extracted using the
online XRT products building
pipeline\footnote{\url{http://www.swift.ac.uk/user\_objects/}} 
\citep{evan09,goad07}. 
We used the XRT specific tasks {\it XSELECT}, {\it XIMAGE} and {\it SOSTA} of the package 
HEAsoft\footnote{\url{http://heasarc.gsfc.nasa.gov/docs/software/lheasoft/}}
to carry out the spectral analysis and obtained  3-$\sigma$ limits on the count rates.

The SN was not detected in any of the X-ray observations.
 The count-rate simulator {\it  WebPIMMS} \footnote{\url{https://heasarc.gsfc.nasa.gov/cgi-bin/Tools/w3pimms/w3pimms.pl}}, using a temperature of 5\,keV and Galactic absorption column density of
$N_H=1.7 \times 10^{20}$\,cm$^{-2}$ towards the SN direction \citep{nh1,nh2}, was used to estimate the  upper limits on the  unabsorbed flux of the SN in  Table \ref{xray}.
A distance of 15 Mpc was used to convert the  fluxes into unabsorbed  luminosities.

\begin{figure}[t]
\centering
	\includegraphics[width=0.69\textwidth]{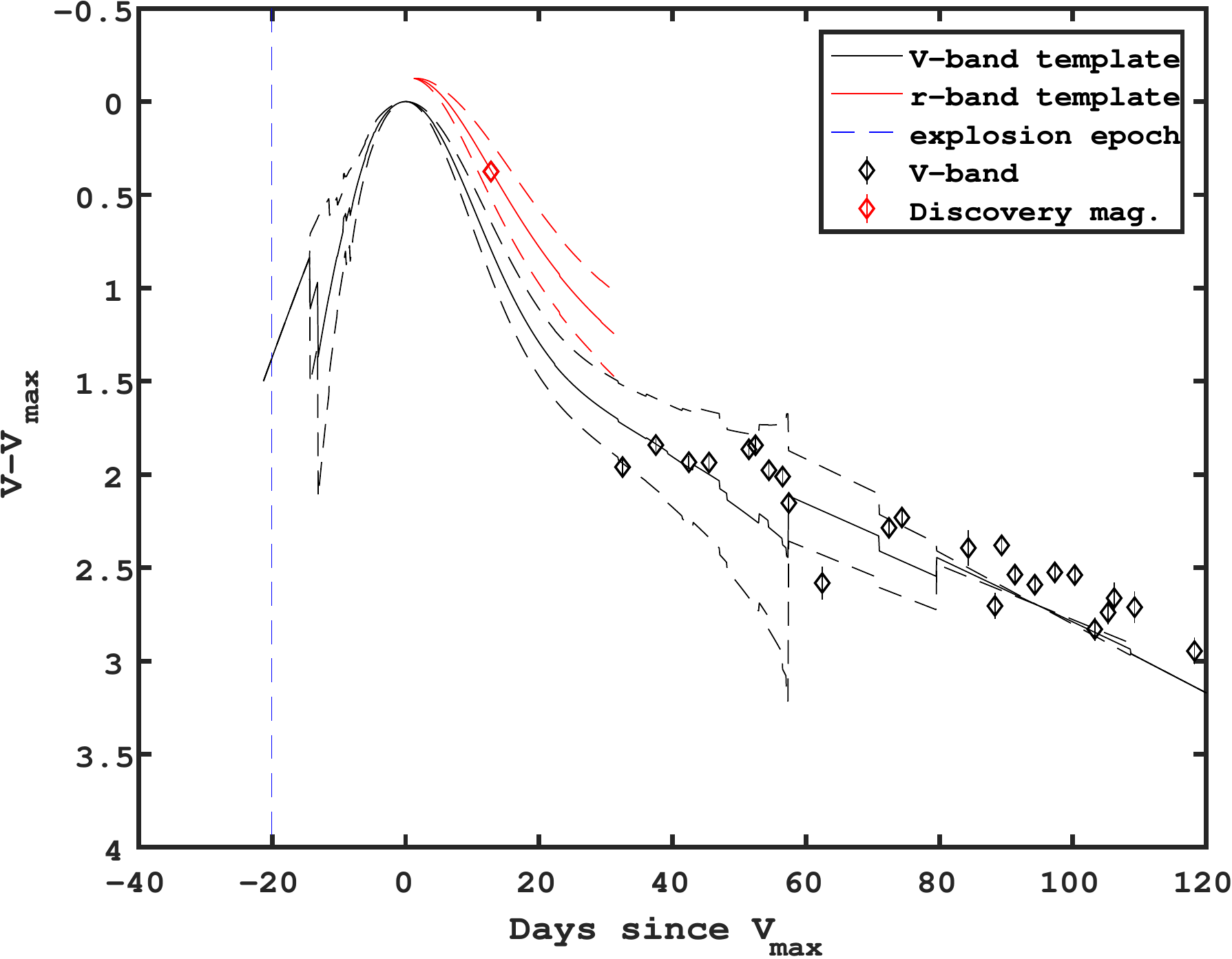}
	\caption{Match between optical light curves and $V$ and $r$-band templates from \citep{taddia2018}. The $V$ band is shown in black and the $r$ band in red. The uncertainty of the templates are marked by dashed curves. The estimated explosion epoch, is marked by a vertical dashed blue line, and was obtained assuming a standard rise-time for SNe Ib from \citet{taddia2018}.}
	\label{taddia}
\end{figure}

\section{Constraining the explosion epoch}\label{sec:explosion}

 There is a large uncertainty in the date of explosion for SN\,J1204. The HCT  spectrum on 2014 Oct 29.0 UT best matched with   several normal type-Ib SNe a few weeks after the maximum \citep{srivastav2014}. Since the typical rest-frame rise time of Type Ib SNe is $\sim 21$ days \citep{taddia+15}, the explosion date is likely to be 30--50 days before the discovery. 
 
We attempt to estimate the explosion epoch based on the optical light curves. Our data are mainly in the $V$-band from  the ASAS-SN observations, but we also have an 
unfiltered magnitude point from the discovery telegram \citep[13.9 mag at JD 2456959.37;][]{gress2014} obtained by the MASTER Global Robotic Net 
\citep{lipunov2010}. The discovery magnitude was significantly brighter than the rest of the data, suggesting that the SN was discovered around the peak  
and that the rest of the data belong to the linear tail of the light curve. To quantify the phase of our observations with respect to the $V$-band peak 
magnitude, we compared them to the stripped-envelope SN light curve templates from \citet{taddia2018}, in the $V$- as well as in the  $r$- bands. 
The match between our data and the templates is shown in Fig. \ref{taddia}. Here we considered that on average stripped-envelope SNe peak 0.124 mag
 brighter in the  $r$-band than in the  $V$-band, and that the peak in the $r$-band occurs 1.23 days after the one in  the $V$-band \citep{taddia2018}. We also assumed 
 that the open-filter magnitude from MASTER can be treated as a $r$-band point \citep[see e.g.,][]{tsvetkov2017}. Finally, we only considered the first four $V$-
 band points for the fit, which are those within 50 days since maximum, when the $V$-band template has  a relatively small spread. From the best match 
 between the data and the templates we derive the $V$-band peak epoch (JD 2456946.55$\pm$3.00). From this epoch we subtracted the average $V$-band 
 rise-time for SNe Ib \citep[from ][]{taddia+15}, which is
20.07$\pm$1.86 d. This gives the
best explosion date to be  JD 2456926.5$\pm$3.5.  This corresponds to UT date of
2014 September 26 within an uncertainty of 3.5 days.
This error on the explosion epoch takes into account the typical
light curve shape of stripped envelope SNe. In the unlikely case that SN\,J1204 is similar to SN 2005bf or 2011bm, i.e., to  rare stripped envelope SNe with long rise times,
then our explosion epoch would have happened between 10 and 20 days earlier.

We note that for the computed $V$-band peak epoch, the classification spectrum \citep[obtained on October 29th 2014, ][]{srivastav2014} occurred  $\sim$13 days after the
peak, consistent with the phase indicated by them (a few weeks after maximum). Furthermore, the Helium-I velocity of 8300 km\,s$^{-1}$ reported by \citet{srivastav2014} is in line with the average Helium-I velocities measured for SNe Ib at that phase after peak \citep{taddia2018}.  Thus in this paper, we assume 
2014 September 26 UT as the date of explosion for SN\,J1204.

\section{Modeling and results}
\label{sec:model}

\subsection{Visual inspection of the data}\label{sec:visual}

\begin{figure*}
	\begin{centering}
		\includegraphics[width=0.48\textwidth]{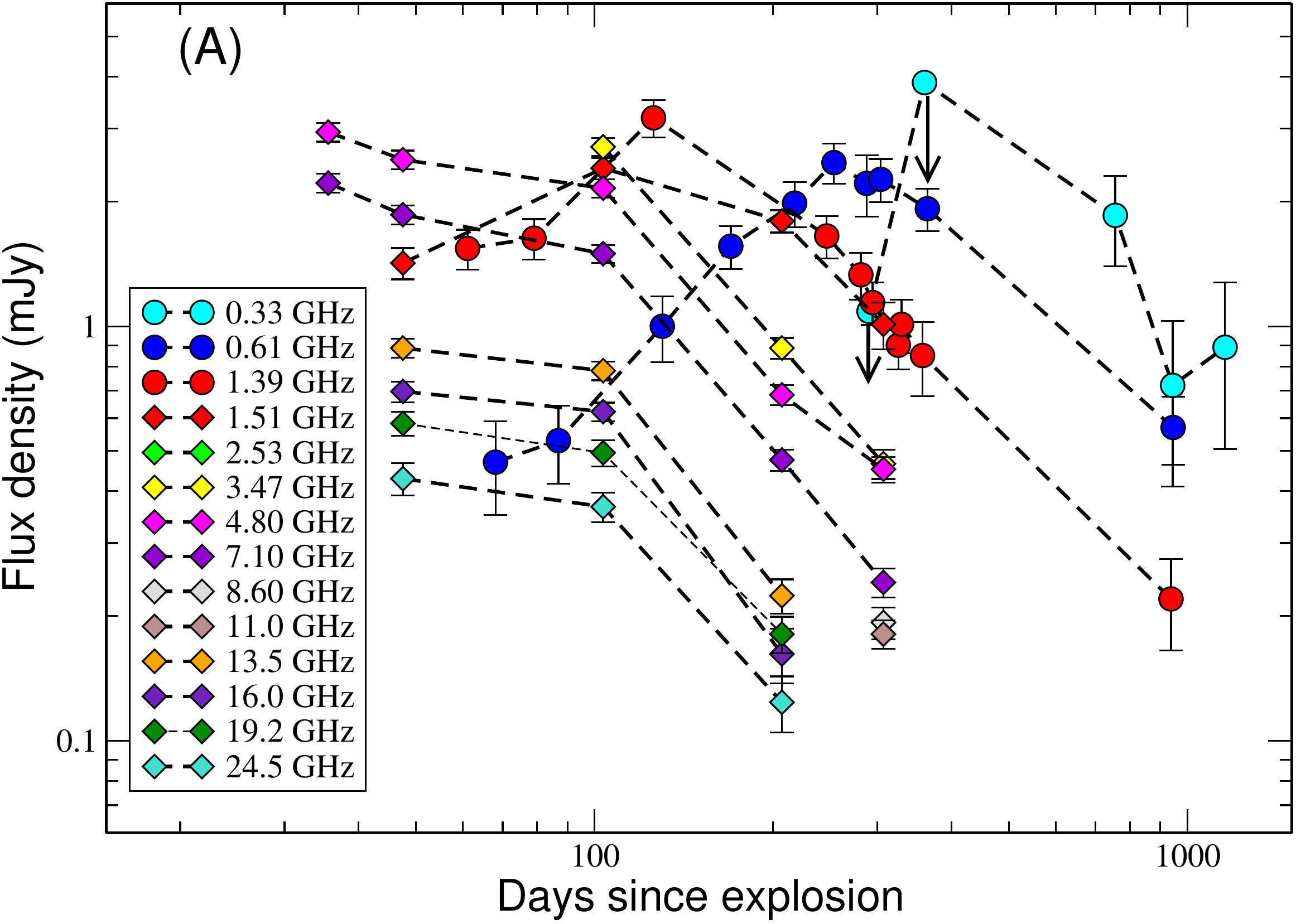}
		\includegraphics[width=0.48\textwidth]{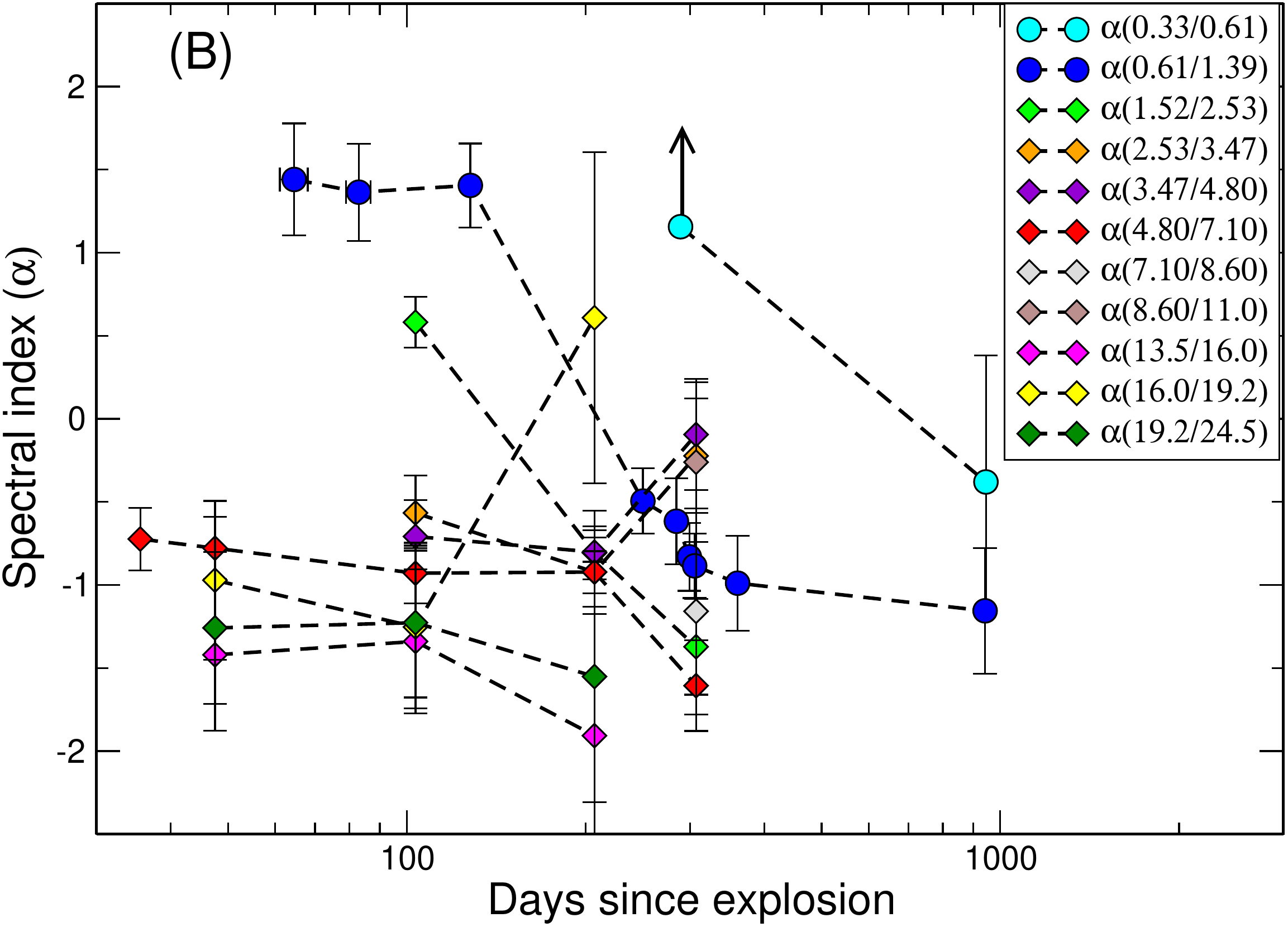}
		\caption{Light curves of SN\,J1204 at frequencies 0.610, 1.4, 2.529, 3.469, 4.799, 7.099, 13.499, 15.999, 19.199 and 24.499 GHz are plotted in the left panel (A). 
		All GMRT data points are represented using circle symbols and all VLA data points are represented using diamond symbols. 
		The L band data ($\sim$1.4 GHz) denoted in red color include both 1.39 GHz GMRT measurements (circles) and 1.515 GHz VLA measurements (diamonds).  We plot the  near-simultaneous spectral indices 
		between adjacent frequencies at various epochs in the right panel (B).}
		\label{full-lc}
	\end{centering}
\end{figure*}

SN\,J1204 is one of the handful  SNe for which extensive data exist at radio  frequencies. 
In Fig. \ref{full-lc}(A), we plot all the radio data from 0.33\,GHz to 24.5\,GHz frequency bands covering the epochs $\sim$35 days till $\sim$1158 days. 
In addition, we plot near-simultaneous spectral indices at various adjacent frequencies   and various epochs in Fig. \ref{full-lc}(B).

The first look at the data reveal that  the  
data are optically thin at frequencies higher than 2.53 GHz onwards.
A significant fraction of low-frequency ($\le 1.5$\,GHz) data  are in optically thick phase.    The first two observations at 0.33\,GHz resulted in upper limits prohibiting us from constraining the
absorption peak at this band, however, we are able to constrain it at other two GMRT frequencies.
The 1.39 and 0.61 GHz light curves peak on  $\sim 126$ d (peak flux density $\sim 3.2$ mJy) and  $\sim 253$ d (peak flux density $\sim 2.5$ mJy), respectively. 
The  peak spectral luminosity at 1.39\,GHz is 
$\sim 8.6\times10^{26}$ erg\,s$^{-1}$\,Hz$^{-1}$, consistent with the radio spectral luminosity of SNe Ib/c which spans a range of $L_{\nu} = 10^{25}-10^{29}$  erg\,s$^{-1}$\,Hz$^{-1}$ \citep{soderberg2007}.

\subsection{Standard model of radio emission}\label{sec:model2}

 In a SN explosion, stellar ejecta are thrown out  into the CSM at supersonic velocities, which  drive a `forward' shock propagating into the CSM and a `reverse' shock moving back into the ejecta. Electrons are accelerated to relativistic energies in the forward shock via diffusive shock acceleration and  produce synchrotron radiation in the presence of amplified magnetic fields. A model for  the ejecta-CSM interaction and its evolution was  developed 
 by \citet{chevalier1982a}. This model follows self-similar solution 
 with the shock radius evolving as power-law in time, i.e.
 $R\propto t^m$, where $m$ is the
 shock deceleration parameter  which is connected to the outer ejecta density profile   $n$ (in $\rho_{\rm ej} \propto R^{-n}$) as
$m=(n-3)/(n-s)$. Here $s$ is  index of the unshocked CSM density profile,  $\rho_{w} $, created by the stellar wind of the progenitor star ($\rho_{w} \propto R^{-s}$).  For a steady stellar wind, $s=2$.
 
 \begin{figure*}
\centering
\includegraphics[width=0.9\textwidth]{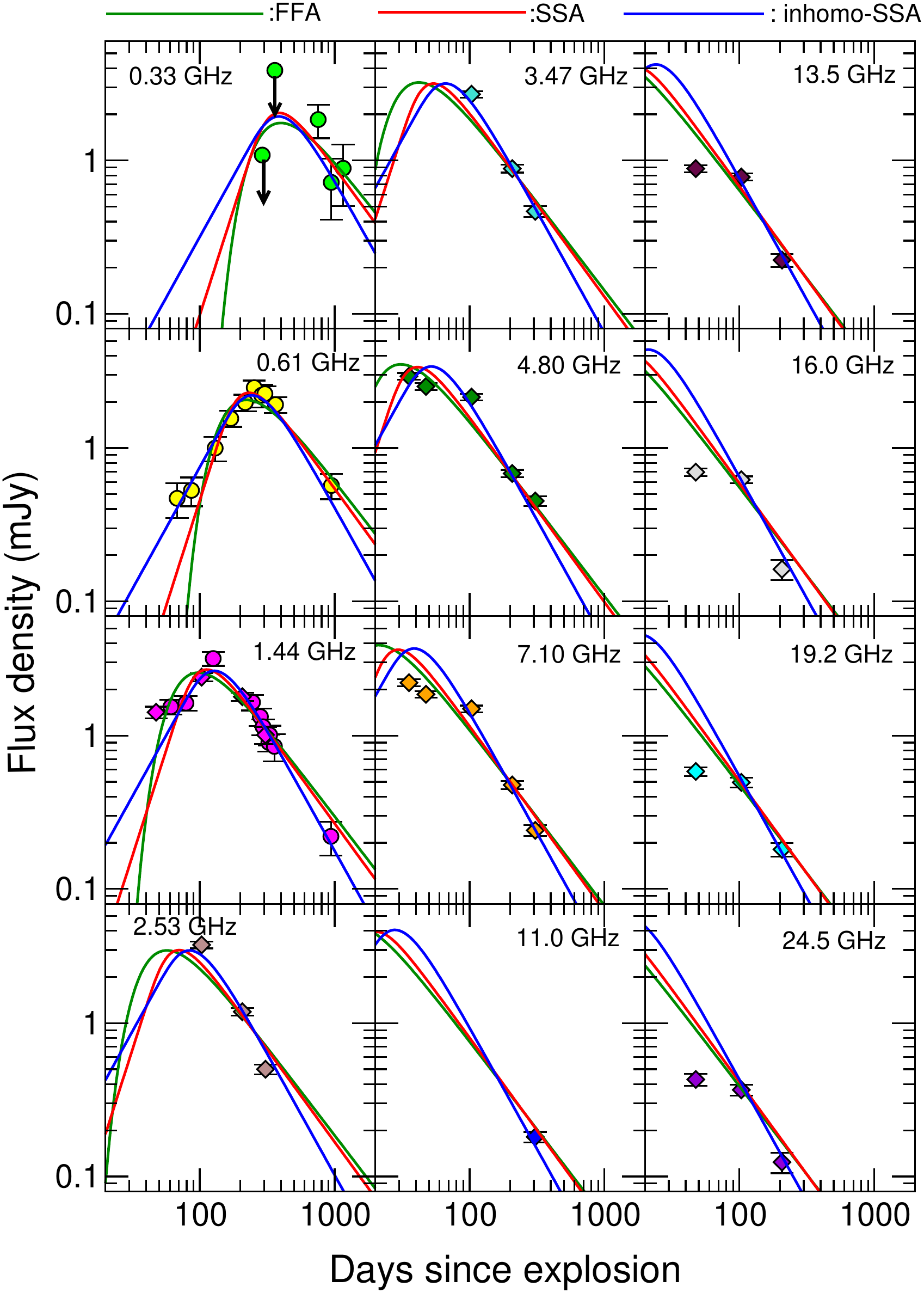}
\caption{ FFA and SSA model fits to the radio light curves of SN\,J1204 at various radio frequencies. The L band data ($\sim$1.4 GHz) includes both 1.39 GHz GMRT measurements (circles) and 1.515 GHz VLA measurements (diamonds). All GMRT data points are represented using circle symbol and all VLA data points are represented using diamond symbol. The green line indicates FFA fits and red line indicate SSA fits. The blue solid lines are the best fit 
curves for the inhomogeneous SSA model.}
    \label{globallcfitsgmrt}
\end{figure*}

 \begin{figure*}
 \includegraphics[width=0.9\textwidth]{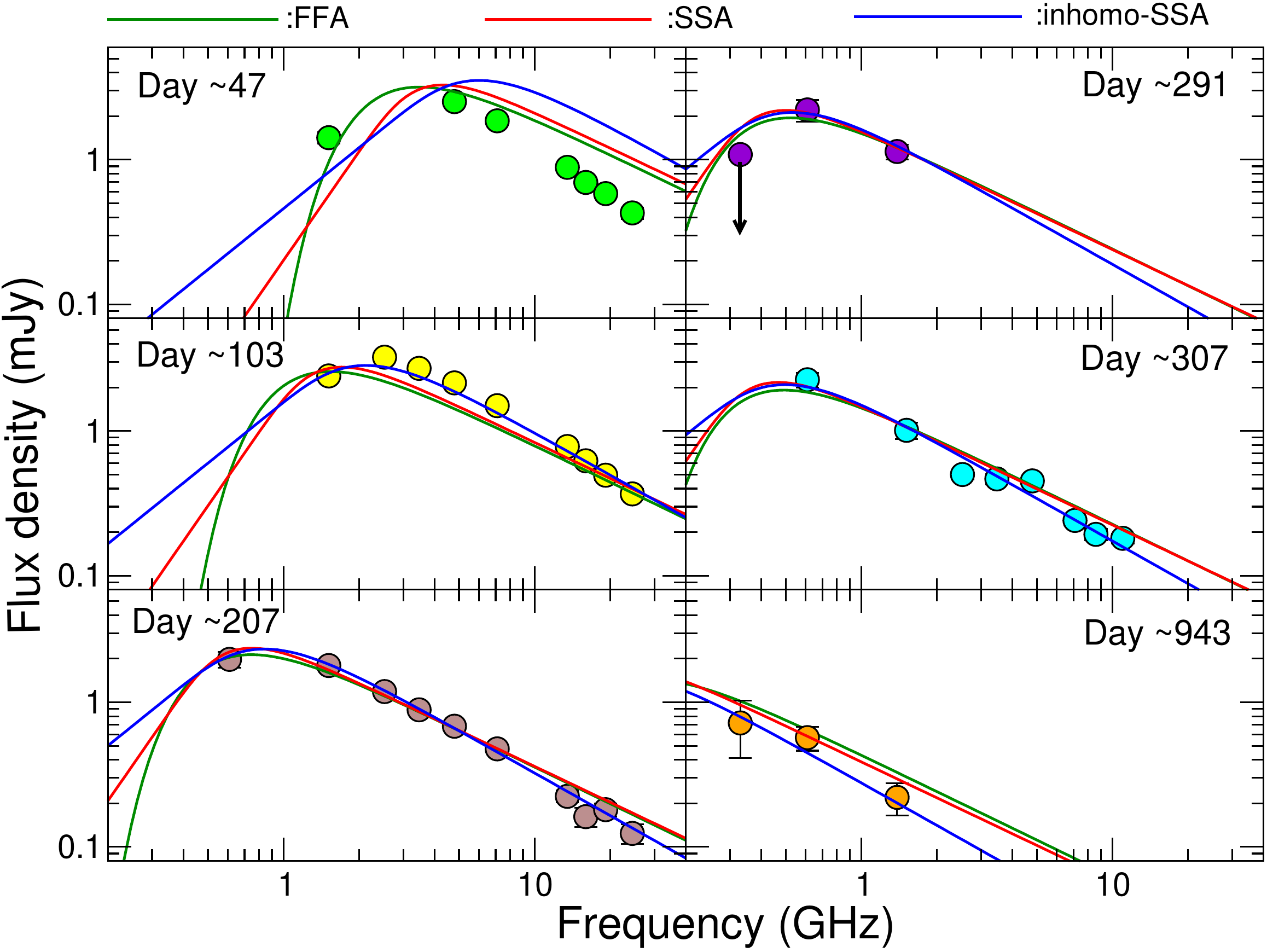}
\caption{Standard FFA and SSA fits to the near simultaneous radio spectra of SN\,J1204 at various epochs post explosion. The model is fitted with the complete data set including both JVLA and GMRT data. The green line indicates FFA fits and red line indicate SSA fits. The blue solid lines are the best fit 
curves for the inhomogeneous SSA model.}
    \label{globalspectrafits}
\end{figure*}

The  radio emission in SNe  
 can be  initially suppressed either due to free-free absorption (FFA) by the ionised CSM \citep{chevalier1982a} or/and
 due to synchrotron self-absorption (SSA) by the same electron population
responsible for the radio  emission  \citep{chevalier1998}. These processes can be distinguished via their early radio light curves and spectra  in the optically thick phase.  As the shell expands, the optical depth decreases and  at optical depth unity SNe  spectra
show  transition from optically thick to optically thin regime indicated by a  change in the sign of spectral index.
Observations spanning over this transition are critically important to pin down the dominant absorption processes, which give information about the 
magnetic field, size of the emitting object, density, mass-loss rates  etc.

 Even though SNe Ib/c generally have SSA as the dominant absorption mechanism,  we start our modeling considering  both FFA and SSA mechanisms. 

In a model where  FFA is the dominant absorption mechanism, the radio flux density, $F(\nu,t)$, can be expressed  as  \citep{weiler2002}

\begin{equation}
F(\nu,t)=K_{1} \left( \frac{\nu}{1\hspace{0.1 cm} \rm GHz}\right)^{-\alpha} \left( \frac{t}{100\hspace{0.1 cm} \rm d}\right)^{-\beta}e^{-\tau_{\rm ffa} (\nu, t)}
\label{eq:ffa1}
\end{equation}
where K$_{1}$ is a normalization factor whose value will be equal to the radio flux density at 1\,GHz measured on day 100 after the explosion. The parameters $\alpha$ and $\beta$ denote the spectral and temporal indices, respectively, in the optically thin phase. The radio spectral index $\alpha$ can be related to the electron energy index $p$ (in $N(E)\propto E^{-p}$) as $\alpha=(p-1)/2$. 
Here $\tau_{\rm ffa}$ is the optical depth  characterized by the free-free absorption due to the ionized CSM external to the emitting material, and can be written as 
 
\begin{equation}
\tau_{\rm ffa}(\nu,t)=K_{2}\left(\frac{\nu}{1\hspace{0.1 cm}\rm GHz}\right)^{-2.1}\left(\frac{t}{100\hspace{0.1 cm} \rm  d}\right)^{-\delta}
\label{eq:ffa2}
\end{equation}
where $K_{2}$ denotes the free-free optical depth at 1 GHz measured on day 100 after the explosion. As the blast wave expands, the optical depth decreases with time as $t^{-\delta}$, where the 
index of optical depth evolution $\delta$ is related to shock deceleration parameter $m$ as  $m=\delta/3$. 

{\color{blue}
\begin{deluxetable*}{ccccc}
\tablecaption{Best fit   parameters values  for the  various models to the radio data \label{fittedpara-gmrt} }
\tablehead{
 \colhead{FFA} & \colhead{SSA} &  \colhead{SSA} &
 \colhead{SSA-inhomo} &  \colhead{SSA-inhomo}\\
\colhead{(full data)} &\colhead{(full data)} &  \colhead{(610 MHz)}  &\colhead{(full data)}& \colhead{($t >87$ d)}
}
\colnumbers
\startdata
$K_{1}=5.75\pm0.95$ &          $K_1=1.56\pm0.29$   & $K_1=0.70\pm0.04$
& $K_1=1.50\pm0.14$ & $K_1=1.45\pm0.13$ \\
$K_{2}=   1.07 \pm0.21$ &        $K_2=3.72\pm1.06$ & $K_2=28.55\pm 5.90$
&  $K_2=6.56\pm1.41$  &  $K_2=8.00\pm1.00$ \\
$\alpha=0.84 \pm 0.12$ &      $\beta'=2.76\pm0.27$ &      $\beta'=1.49\pm0.12$
& $\beta'=1.59 \pm0.12$ & $\beta'=1.56 \pm0.21$\\
$\beta=1.15 \pm 0.08$ &       $\beta=1.21\pm0.09$   & $\beta=1.59\pm0.12$ 
& $\beta=1.59\pm0.10$ & $\beta=1.72\pm0.05$\\
$\delta=2.20 \pm 0.24$ &      $p=2.65\pm0.20$ & $\cdots$
 & $p=2.98\pm0.18$  & $p=3.04\pm0.05$  \\
$\chi_{\mu}^{2}=8.45$ &             $\chi_{\mu}^{2}=7.73$  &             $\chi_{\mu}^{2}=0.96$
& $\chi_{\mu}^{2}=4.36$  & $\chi_{\mu}^{2}=1.62$ \\
{\it d.o.f.}=52 & {\it d.o.f.}=52 & {\it d.o.f.}=6 & {\it d.o.f.}=52 &  {\it d.o.f.}=39\\
\enddata
\tablecomments{Here the parameters are described in \S \ref{sec:model2} and \S \ref{sec:inhomo}, and $\chi_{\mu}^{2}$ is the
reduced Chi-Square, and {\it d.o.f.} is the degrees of freedom. }
\tablecomments{The SSA model assumes model 1 of \cite{chevalier1996} with the magnetic energy density and relativistic electron energy density scale with the post-shock energy density.}
\end{deluxetable*}
}

For the SSA dominated synchrotron emission from SNe,  the radio flux density can be written as \citep{chevalier1998}:
\begin{equation}
F(\nu,t)=K_{1} \left( \frac{\nu}{1\hspace{0.1 cm} \rm GHz}\right)^{2.5} \left( \frac{t}{100\hspace{0.1 cm} \rm d}\right)^{\beta'} (1- e^{-\tau_{\rm ssa}(\nu, t)})
\label{eq:ssa1}
\end{equation}
where the optical depth is characterised by SSA due to the relativistic electrons at the forward shock. The SSA optical depth $\tau_{\rm ssa}$ is given by
\begin{equation}
\tau_{\rm ssa}(\nu, t)=K_{2}\left(\frac{\nu}{\rm 1\hspace{0.1 cm} GHz}\right)^{-(\alpha+2.5)}\left(\frac{t}{100\hspace{0.1 cm}  \rm d}\right)^{-(\beta'+\beta)}
\label{eq:ssa2}
\end{equation}
K$_{1}$ and K$_{2}$ are the flux density and optical depth normalisation factors similar to the case of FFA. The flux density in the optically thick and
thin phases evolve as $\nu^{2.5}$ and $\nu^{-\alpha}$, respectively. Here spectral index $\alpha$ is related to electron energy 
index $p$ as $\alpha=(p-1)/2$. Similar to FFA model, $t^{-\beta}$ is the time evolution of flux density in the optically  thin phase, whereas,
 $t^{\beta'}$  is the flux density evolution with time in the optically thick  phase. While $\beta'$ depends on 
shock deceleration parameter $m$, $\beta$ depends on $m$ as well as electron energy index $p$. The exact form depends upon the 
scalings of magnetic field, $B$,  and the electron energy density \citep{chevalier1996}. For example, if magnetic energy density and 
relativistic electron energy density scale with post shock energy density \citep[model 1 of ][]{chevalier1996}, then 
optically thick light curve
$F_\nu(t) \propto R^2B^{-1/2}$ will lead to  
$\beta'=2m+0.5$ and then $\beta=(p+5-6m)/2$. This involves an  assumption that the magnetic field was built up by turbulent motions. However, if compression of the CSM magnetic field determines the relevant magnetic field  scaling and relativistic electron energy
density scales with the flux of particles into the shock front,  then $\beta'=5m/2$ and $\beta=(p-1)m/2$ \citep[model 4 of ][]{chevalier1996}. 
 When there are substantial data in the optically thick phase,  the parameters $\beta'$, $\beta$ and $p$ can be independently obtained, and one can determine the relevant scalings in a  particular case.

 We now model the full data using both the FFA and the SSA models.
We first  perform the FFA model fit keeping $K_{1}$, $K_{2}$, $\alpha$, $\beta$ and $\delta$ as free parameters. The best-fit parameters are obtained using the $\chi^{2}$
 minimisation. 
The fitted models along with the measured spectra and light curves are shown in  
Figs. \ref{globallcfitsgmrt} and \ref{globalspectrafits}. The best-fit parameters are given in  the Table \ref{fittedpara-gmrt}. 
We note that the  reduced chi-square ($\chi_{\nu}^2$) for the 
FFA model is quite high ($\chi_{\nu}^2=8.45$),  suggesting that the FFA model is not a good fit to the radio data. 

We now fit the data with the SSA model as given in Eqns. \ref{eq:ssa1}
and \ref{eq:ssa2}. The 
best-fit parameters for the SSA model are  given in  Table \ref{fittedpara-gmrt} and plotted in Figs  \ref{globallcfitsgmrt} and \ref{globalspectrafits}. The reduced chi-square is this case is 
$\chi_{\mu}^{2}=7.73$. While the SSA model performs slightly better  than the FFA model, it fails to provide  an acceptable fit to the data.

\section{A need for a non-standard model}\label{sec:non-standard}

Unfortunately neither the  FFA  nor the SSA models  provide a good fit to the radio data of SN J\,1204. Visual inspection of Figs.   \ref{globallcfitsgmrt}  and \ref{globalspectrafits}  suggests that the data $\le 103 $ days  are not fit well with the standard models.   The model  over-predict the flux on $\sim47$ d and under-predict
the flux at $\sim 103$ d. 
The low frequency light curves show flattening at the early epochs in the optically thick phase. 
The discrepancies between data and models are more pronounced at earlier epochs.

\begin{figure}
 \centering
 	\includegraphics*[width=0.69\textwidth]{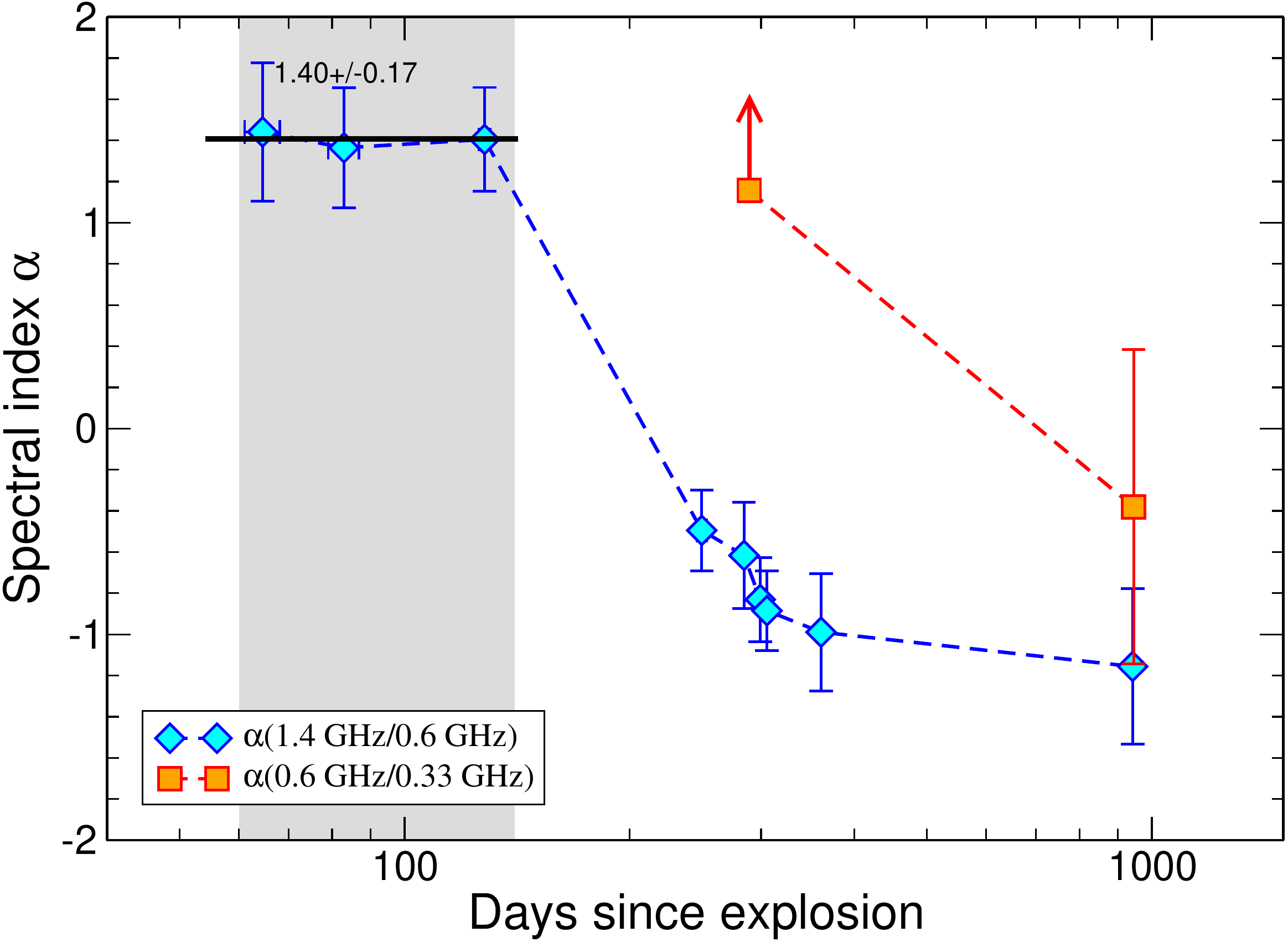}
 	\caption{Spectral index evolution between 1390 and 610 MHz (blue diamonds) and between 610 and 325 MHz bands (orange squares). }
 	\label{fig:spec}
 \end{figure}

To understand the early time behaviour of the radio data, we investigate the estimated spectral indices  in the optically thick phase, which are mainly at the   GMRT frequencies. 
We plot them  in Fig. \ref{fig:spec}  and tabulate
in Table \ref{tab:specind}.  We notice that during the  first three epochs, i.e. between days  64 to 127,  the spectral indices are $\alpha(1390/610) \sim 1.4$. 
This value is 
much flatter than the spectral index 2.5 expected in the SSA model or steeper value  in the FFA model. The values of  $\alpha(610/325)\ge1.16$ is a lower limit on day $\sim$289, which is also consistent with the above values.

While the spectral indices are expected to flatten near the spectral peaks, the 1390 MHz peak occurs at day 126 and 610 MHz peak much later (see \S
\ref{sec:visual}). Thus only the last data point around 127 d may be affected by this effect. The fact that the spectral indices have flatter values from 64 d onwards, indicates that this is a real effect.  The  average value of the  optically thick spectral index from the first three epochs is $1.40\pm0.17$
(Fig. \ref{fig:spec}).

\begin{deluxetable}{ccc}
	\tablecaption{Spectral indices for near simultaneous measurements at the GMRT frequencies \label{tab:specind} }
	\tablehead{
		\colhead{Age\tablenotemark{a} } & \colhead{Spectral index} & \colhead{Spectral index}\\
		\colhead{Days} & \colhead{$\alpha_{1390/610 \rm \,MHz}$} &  \colhead{$\alpha_{610/325\rm \,MHz}$}
	}
	\startdata
	$ 64.57\pm 3.48$ &  $ 1.44\pm 0.34$ & $\cdots$\\
	$ 82.98\pm 3.94$ &  $ 1.36\pm 0.29$ & $\cdots$\\
	$ 127.92\pm 2.17$ &  $ 1.40\pm 0.25$ & $\cdots$\\
	$ 249.92\pm 3.50$ &  $ -0.49\pm 0.20$ & $\cdots$\\
	$ 284.58\pm 3.15$ &  $ -0.62\pm 0.26$ & $\cdots$\\
	$ 289.12\pm 1.39$ & $\cdots$ &  $ \ge 1.16$ \\
	$ 299.18\pm 4.64$ &  $ -0.83\pm 0.20$ & $\cdots$\\
	$ 305.46\pm 1.64$ & $ -0.89\pm 0.20$\tablenotemark{b} & $\cdots$\\
	$ 360.89\pm 3.44$ &  $ -0.99\pm $ 0.29 & $\cdots$\\
	$ 942.55\pm 4.00$ &  $ -1.16\pm 0.37 $ & $\cdots$\\
	$ 945.54\pm 1.01$ & $\cdots$ &  $ -0.76\pm0.69$ \\
	\enddata
	\tablenotetext{a}{The age is calculated assuming 2014 September 26 (UT) as the date of explosion. The  range in the epochs reflect the time span of the near-simultaneous measurements.}
	\tablenotetext{b} {With VLA 1.5 GHz measurement,}
\end{deluxetable}

To probe the nature of absorption further, we look  at the radio  light curve at 610 MHz,  the best sampled 
 frequency in the optically thick phase. 
We again fit  the standard FFA and SSA models, using Eqns. \ref{ffa:lc} and \ref{eq:610}, respectively.

\begin{equation}
F(t)_{[\rm FFA]}=K_{1} \left( \frac{t}{100\hspace{0.1 cm} \rm d}\right)^{-\beta}  \left[\exp{\left(-K_{2}\left(\frac{t}{100\hspace{0.1 cm} \rm  d}\right)^{-\delta}\right)}\right]
\label{ffa:lc}
\end{equation}

\begin{equation}
F(t)_{[\rm SSA]}=K_{1} \left( \frac{t}{100\hspace{0.1 cm} \rm d}\right)^{\beta'} \left[1- \exp\left({-K_{2}\left(\frac{t}{100\hspace{0.1 cm}  \rm d}\right)^{-(\beta'+\beta)}}\right)\right]
\label{eq:610}
\end{equation}

\begin{figure}
\centering
    \includegraphics[width=0.69\textwidth]{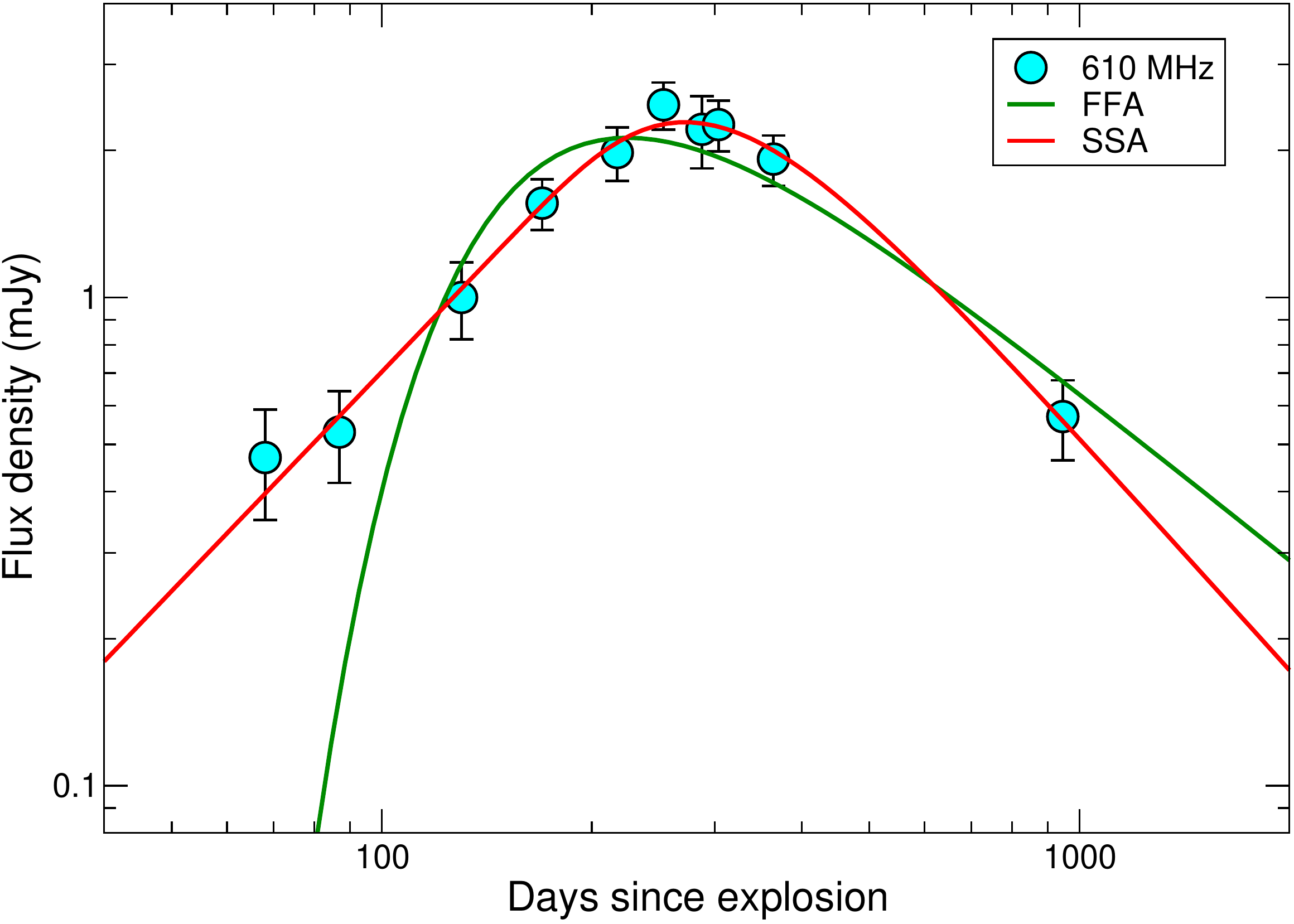}
\caption{SSA model  fit to GMRT 610 MHz light curve. The data is  dominated by the points in the optically thick part of the light curve. }
    \label{lc-fit610-only}
\end{figure}

While the FFA  model  gives a  poor fit, SSA model seems to fit the data reasonably well (Fig. \ref{lc-fit610-only}). 
However, the
best fit values of $\beta'=1.49\pm0.12$ and $\beta=1.59\pm0.12$ obtained in these fits (column 3 of Table \ref{fittedpara-gmrt}) will imply   $m=0.55\pm0.06$ and $p=1.45\pm0.43$. The value of $p$ is
unrealistically small and is inconsistent with the observations  (Fig. \ref{full-lc}). 
  The derived value  of $m$ is much smaller than  typical values seen in Type Ib SNe and will indicate very high deceleration at this young age, which is unlikely.   
Hence we conclude that despite providing acceptable fit,  the standard SSA model does not represent a physically viable model  to 610\,MHz light curve.

Unphysical values of model parameters obtained above combined with the much flatter spectral index in the optically thin phase, $\alpha'=1.4\pm0.17$, indicate that the standard homogeneous SSA emission model  does not fit the radio emission in SN\,J1204.

\subsection{An inhomogeneous model}\label{sec:inhomo}

\citet{bjornsson13} and \citet{bk17} have explained the flatter optically thick evolution of the radio spectra in terms of  inhomogeneities in the radio emitting region caused by the variations in the distribution of magnetic fields and relativistic electrons.   \citet{bk17}  have shown that if the emission structure is inhomogeneous, then  fitting a standard homogeneous model to observations around the peak frequency gives a lower limit to the source radius.
\citet{bjornsson13} and \citet{bk17}  have quantified the  variation of the magnetic field over the projected source surface  by a source covering factor,  $f_{B,\rm cov}$.
As per their formulation, even though the locally emitted spectrum is that of standard synchrotron, the $f_{B,\rm cov}$ will give rise to a range of optical depths over the source,  broadening the
observed spectrum. 
Since the covering factor is maximum at  frequencies substantially below that of the spectral peak \citep{bk17}, the  detailed observations at low frequencies are best to probe the
inhomogeneities.

In \S \ref{sec:a1}, we develop an inhomogeneous emission model adopted from \citet{bk17}.
 As per this formulation, the inhomogeneous model will alter the spectra for magnetic fields ranging between $B_0<B<B_1$ (Fig. \ref{fig:appendix}), and radio emission will follow the  standard
 homogeneous SSA formulation at frequencies corresponding to magnetic fields outside this range, i.e.
 \begin{equation}
F(\nu)\propto \begin{cases}
\nu^\frac{5}{2},
     & \nu < \nu_{\rm abs}(B_0)    \\
   \nu^{\frac{3p+7+5\delta'-a(p+4)}{p+2(1+\delta')}},
     &  \nu_{\rm abs}(B_0)< \nu < \nu_{\rm abs}(B_1)\\
     \nu^\frac{-(p-1)}{2},
      & \nu > \nu_{\rm abs}(B_1)  \: , 
\end{cases}
\label{eq:cases}
\end{equation}
Here $ \nu_{\rm abs}$ is the SSA frequency, $a$ is defined as $P(B) \propto B^{-a}$, where  $P(B)$ is the probability
of finding a particular value of $B$ within $B$ and $B+dB$ (Eq. \ref{eq:a1}). We define $\alpha'$ as spectral index in the  SSA optically thick phase in the inhomogeneous model.
Thus in our case $\alpha' \equiv (3p+7+5\delta'-a(p+4))/(p+2(1+\delta'))=1.4$.
Here  $\delta'$ indicates a correlation between  the distribution of relativistic 
electrons with the distribution of magnetic field strengths (see \S \ref{sec:a1}).
For $\delta'=0$ the inhomogeneities  in the relativistic electrons distribution are not correlated with the inhomogeneities  in the
magnetic field. 
For $\delta'=1$,  the inhomogeneities between the two distributions are correlated. 

Since our radio  observations in the optically thick phase indicate $\alpha'<5/2 $ at all epochs, this suggests we are in the regime  $ \nu > \nu_{\rm abs}(B_0) $  for  observed frequency range.
For  the observed $F_\nu \sim \nu^{1.4}$ in the optically thick phase,   and for $p=3$, we obtain $a=1.3$
 for $\delta'=0$ and $a=1.6$ for $\delta'=1$.

In order to take the inhomogeneities into account, we use a model, in which  optically thick spectral index follows  $\alpha'=1.4$ in
Eqns. \ref{eq:ssa1} and \ref{eq:ssa2} :
\begin{equation}
F(\nu,t)=K_{1} \left( \frac{\nu}{1\hspace{0.1 cm} \rm GHz}\right)^{\alpha'} \left( \frac{t}{100\hspace{0.1 cm} \rm day}\right)^{\beta'} (1- e^{-\tau_{\rm ssa}(\nu, t)})
\label{eq:ssa1-inhomo}
\end{equation}
where the optical depth is characterised by the SSA due to the relativistic electrons at the forward shock. The SSA optical depth $\tau_{\rm ssa}$ is given by

\begin{equation}
\tau_{\rm ssa}(\nu, t)=K_{2}\left(\frac{\nu}{\rm 1\hspace{0.1 cm} GHz}\right)^{-(\alpha'+\frac{p-1}{2})}\left(\frac{t}{100\hspace{0.1 cm}  \rm day}\right)^{-(\beta'+\beta)}
\label{eq:ssa2-inhomo}
\end{equation}

The best fit values are tabulated in column 4 of Table \ref{fittedpara-gmrt}. We plot the best fit inhomogeneous model in Figs. 
 \ref{globallcfitsgmrt} and \ref{globalspectrafits}. While the inhomogeneous model fits are  better than the  standard SSA and FFA fits,
  and  $\chi_\nu^2$ improved by nearly a factor of $\sim 2$,  the early data still deviates from the inhomogeneous SSA model.
Fig. \ref{globalspectrafits} indicates that the model spectrum on day $\sim47$ does not represent the data well.The model light curves at early epochs are  also discrepant with the data (Fig \ref{globallcfitsgmrt} ).

This suggests that a global model assuming constant $\beta'$, $\beta$ and $p$
will not fit the data well at all the  epochs.  
This situation can be reconciled if  there is an evolution in the blast-wave parameters with time, likely at early epochs.

\subsection{Shock passing through a shell}\label{sec:shell}

\begin{figure}
\includegraphics[width=0.51\textwidth]{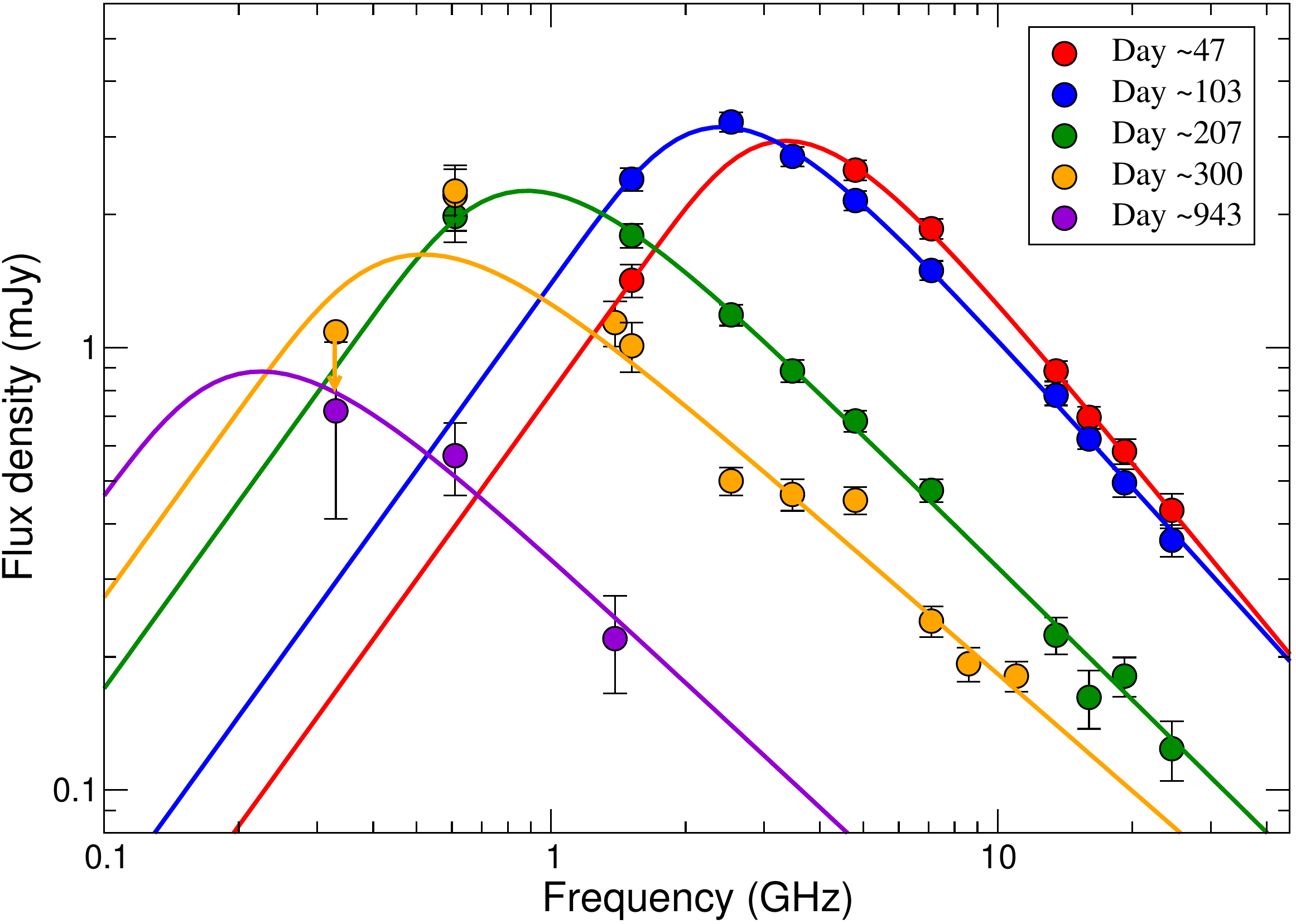}
\includegraphics[width=0.5\textwidth]{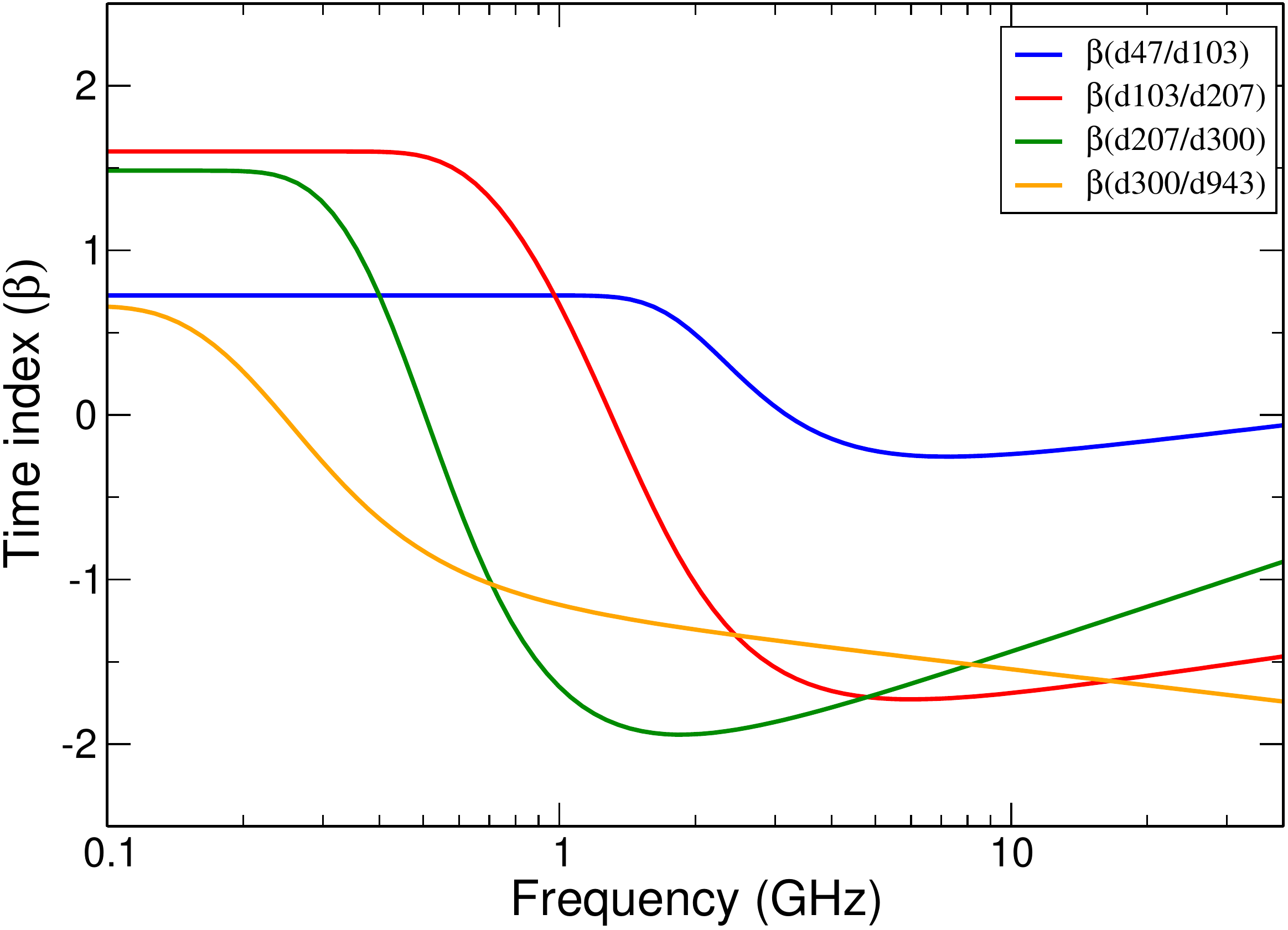}
 \caption{\textit{Top panel:} Inhomogeneous synchrotron self-absorption model fit to individual spectra of SN\,J1204 on Days 47, 103, 207, 300
 and 943, respectively. In the
 bottom panel, we plot the evolution of temporal index $\beta$ as a function of frequency.}
    \label{ssa-spectranblast wave para evolution}
\end{figure}

To understand the dynamical evolution of the blast wave and to investigate the inconsistency with the global fit parameters, 
we study the near-simultaneous spectra of the SN at individual epochs (Fig. \ref{ssa-spectranblast wave para evolution}.)
Since we established  the need for an inhomogeneous model in  \S \ref{sec:inhomo}, we model the spectra of SN\,J1204 at each epoch with  the inhomogeneous SSA model:
\begin{equation}
F(\nu)=1.582\, F_{\rm abs} \left( \frac{\nu}{\nu_{\rm abs}}\right)^{\alpha'} \left[ 1-\exp \left( -\left( \frac{\nu}{\nu_{\rm abs}}\right)^{-(\alpha'+\alpha)} \right)\right]
\label{eq:ssa-spec-inhomo}
\end{equation}
where we use  $\alpha' \equiv1.4$. $F_{\rm abs}$ is the peak flux density at a frequency $\nu_{\rm abs}$ at a given epoch.

 There are a few things to decipher from the individual   spectra.  The spectra evolve very little between days $\sim47$ and $\sim103$
 (Fig. \ref{ssa-spectranblast wave para evolution}). 
 The peak flux density  $F_{\rm abs}$ and peak frequency $\nu_{\rm abs}$  have  a nearly flat evolution   $F_{\rm abs} \propto t^{0.24\pm0.08}$ and  $\nu_{\rm abs} \propto t^{-0.38\pm0.06}$ between these two epochs. 
  In addition, we note that there are 7 and 8 independent data points in spectra on day 47 and 103, respectively, and all of these data are consistent with much slower evolution than expected in the
standard synchrotron emission model.
 This can also be seen in the time evolution plot, where $\beta $ is close to zero at high frequencies during these two epochs. 
 Such a situation may arise  if the shock is crossing through a  shell and has slowed down due to a high density of the shell, causing the time evolution of the parameters  to slow down between these two epochs. 

\begin{figure}
\begin{centering}
\includegraphics[width=0.68\textwidth]{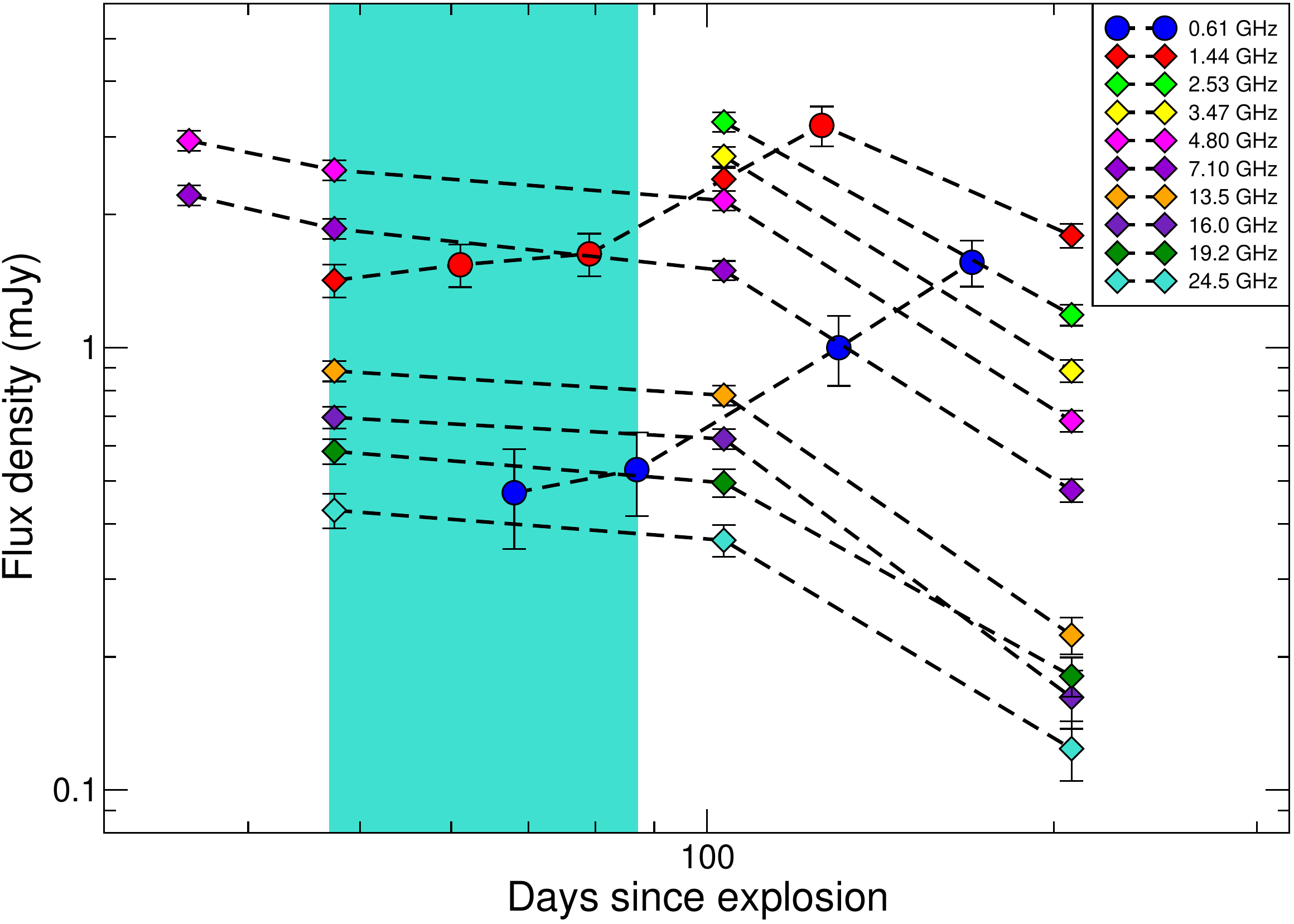}
  \caption{Radio light curves of SN\,J1204 up to day $\sim 200$.}
    \label{zoom}
    \end{centering}
\end{figure}

While it is difficult to decipher the exact  duration for which  the shock is passing though the shell, we attempt to constrain it based on our radio data. 
For this purpose, we re-plot the light curve zooming into early times  (Fig. \ref{zoom}).
Fig. \ref{zoom}  indicates that while the  flux density in the earliest data  between day 35 and  47  evolves as 
 $t_{(4.8\, \rm GHz)}^{-0.53\pm0.25}$ and  $t_{(7.1\, \rm GHz)}^{-0.60\pm0.25}$   at 4.8  and 7.1 GHz bands, respectively, 
 the evolution of the flux density slows down   as
$t_{(4.8\, \rm GHz)}^{-0.20\pm0.09}$ and $t_{(7.1\, \rm GHz)}^{-0.28\pm0.09}$  between days 47 to 103, respectively. This again steepens and evolves as  $t_{(4.8\, \rm GHz)}^{-1.65\pm0.12}$ and $t_{(7.1\, \rm GHz)}^{-1.65\pm0.12}$ post day 110.
 While the relatively flatter evolution between  35 and 47 days at 4.8 and 7.1 GHz bands could be reconciled with a situation where 
we may be witnessing the shock  soon after the peak transition at these frequencies and  tracing the broader top likely due to inhomogeneities,  the significant flattening  between day 47 and 103  cannot be explained in this framework.

This suggests that shock was probably moving into a smooth CSM  at the first epoch $d\sim35$, but entered  into a
higher density shell some time during $d\sim47$. 
The steep evolution between days 103 and 207 suggests that it was out of the shell by the time VLA observations commenced on $d\sim103$. An additional clue 
on the length of time it took for the shock  
to cross the shell
comes from the optically thick data
points at 1.4 and 0.6 GHz bands.  The first three epochs of 1.4 GHz light curves (covering $\sim79$ days), and first two epochs at 0.61 GHz (covering
$\sim 87$ days) are 
 flatter than the other optically thick data points at this frequency (Fig. \ref{zoom}). This indicates that shock likely stayed in
the dense shell up to $t \sim 87$d and was out of the shell afterwards. Thus we infer that the shock likely remained in the shell  during  $47$ to 
$87$ days after the SN explosion.  However, we note that the transition at day $\sim47$ is much less secure than that at day $\sim87$.

 After emerging from the shell, the shock velocity is expected to stay roughly constant \citep{vanmarle2010} or even accelerate \citep{harris+16}. This transition phase may last a few dynamical timescales, i.e., until the swept-up mass starts to dominate the extra mass in the shell. Regardless of the details of this phase,  the time evolution of $\nu_{\rm abs}$ should speed up and approach that pertaining to the time before the shock entered the shell.

  In order to evaluate the possible effects of FFA after the shock entered the shell, we examine  $\tau_{\rm ffa}(\nu_{\rm abs})$, which is the FFA optical depth at $\nu_{\rm abs}$.  As discussed by \citet{bl14}, this is a maximum value for the free-free optical depth $\tau_{\rm ffa}$,  because it is set by the temperature resulting from the actual heating due to the absorbed synchrotron emission itself. 
   Neglecting terms of order unity, $\tau_{\rm ffa}(\nu_{\rm abs})$ is \citep{bl14}
\begin{equation}
	\tau_{\rm ffa}(\nu_{\rm abs}) \sim 3 \left( \frac{v_{\rm sh}}{10^4\rm km\,s^{-1}}\right )^{-\frac{9}{5}}  \left( \frac{t}{10\,\rm d} \right)^{-\frac{12}{5}} \left( \frac {\nu_{\rm abs}}{ 10\,\rm GHz} \right)^{-\frac{13
	}{5}} \left(\frac{\dot{M}_{\rm -5}}{v_{\rm w,1}}\right)^{\frac{7}{5}}
	\label{eq:shell}
\end{equation}
where $v_{\rm sh}$ is the velocity of the forward shock in $\rm km\,s^{-1}$.
 The density of the CSM is $\propto \dot{M}_{\rm -5}/v_{\rm w,1}$, where $\dot{M}_{-5}$ is the steady  mass loss rate of the progenitor 
star in units of $10^{-5}$ solar masses per year and $v_{\rm w,1}$ is the corresponding wind velocity in units of 10\,km\,s$^{-1}$, respectively. Since it is likely that the shock velocity is larger than 
$10^4$  km s$^{-1}$ at $t \sim 47$ d, and $\nu_{\rm abs}= 3.5$\,GHz on this day  (Fig. \ref{ssa-spectranblast wave para evolution}), Eq. \ref{eq:shell} imples
\begin{equation}
	\tau_{\rm ffa}(\nu_{\rm abs}) <1\times \left( \frac{\dot{M}_{-5}} {v_{\rm w,1}} \right)^{\frac{7}{5}}
\end{equation}
Since SN\,J1204 is a Type Ib SN, $v_{\rm w,1} \gg 1$ is expected. Unless the mass-loss rate is unusually large (i.e., $\dot{M}_{\rm -5} \gg 1$), FFA  should be negligible in the wind; for example, a wind velocity of $10^3$\,km\,s$^{-1}$ and $\dot{M}_{\rm -5} = 1$ give $\tau_{\rm ffa}(\nu_{\rm abs}) = 1.6 \times 10^{-3}$. 

Although hydrodynamical simulations are needed to describe the passage of the shock through the shell, 
Eq. \ref{eq:shell} can be used to estimate its FFA. Initially, when the forward shock impacts the shell, the already 
shocked mass between the forward and reverse shocks will act as a piston. The shock velocity in the shell is then given, 
roughly, by momentum conservation, i.e., the shock velocity slows down by a factor $(\rho/\rho_{\rm o})^{1/2}$. Here, $\rho$ is 
the shell density and $\rho_{\rm o}$ the density behind the forward shock at impact. Furthermore, observations show that the 
duration of the shell passage is roughly equal to the time for the forward shock to reach the shell. The width of the shell is then
 $\Delta R \approx R (\rho_{\rm o}/\rho)^{1/2}$. The FFA of the shell can then be obtained by multiplying the RHS 
 of Eq. \ref{eq:shell} with $(\rho/\rho_{\rm o})^{7/5} (\Delta R/R) = (\rho/\rho_{\rm o})^{9/10}$; for example, with $v_{\rm w,1} =10^2$ 
 and $\dot{M}_{\rm -5} = 1$, a FFA optical depth of unity in the shell corresponds to $\rho/\rho_{\rm o} \approx 1.3 \times 10^3$. For a 
 strong shock, the density in the wind is $\rho_{\rm o}/4$, which gives a density contrast between the shell and the wind of $3.2\times 10^2$. 
 Furthermore, the associated slow-down of the shock velocity would be  by a factor $38$.  In the standard model, the values of $\nu_{\rm abs}$ and $F(\nu) $
 are expected to evolve roughly inversely with time (cf. also the analytical fits done above). As shown in Figs. \ref{ssa-spectranblast wave para evolution} and \ref{zoom}, the observed slow-down is 
 substantially less than this. Hence, the optical depth to FFA in the shell should also be substantially below unity at   3.5 GHz. 
 and the density contrast between the shell and the wind is expected to be less than 
 $10^2$. 




\begin{figure}[t]
\begin{centering}
\includegraphics[width=0.68\textwidth]{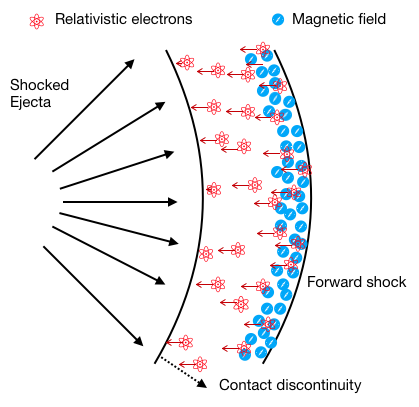}
  \caption{Cartoon diagram of a situation where relativistic electrons are homogeneously  distributed,  but magnetic field is not. 
  The near constant magnetic field is confined within a small distance from the shock front.
  Here one can obtain a case where the main
contributions to both emission and  self-absorption} come from near the shock boundary, and give rise to flat  light curves seen at the VLA frequencies.
    \label{cartoon}
    \end{centering}
\end{figure}

There is another puzzle. If the shock is passing through a dense shell, one would expect the optically thin emission to increase 
 due to the continuous injection of relativistic electrons.
But Fig. \ref{zoom} clearly shows that the  flux evolution at 4.8 GHz and higher frequenies is flatter and optically thin. This could have been explained by cooling, but 
Fig. \ref{full-lc}(B) indicates  that  the optically thin spectral indices are mostly consistent with near constant values and 
do not show any particular steepening in this duration. Thus cooling is unlikely to be the reason for flatter
optically thin light curves. 
  \citet{bjornsson13}  has  explained this situation where a constant magnetic field is confined within a small distance from the shock front in the inhomogeneous model 
(their Eq. 4).  In such a case, relativistic electrons are continuously entering the shocked shell and cascading downwards, hence leaving the width of the radio emitting region almost constant.
Thus the main
contributions to both emission and absorption come from near the shock boundary, and give rise to flat light curves  
 seen at the VLA frequencies. This would then be consistent with
a situation where the inhomogeneities are confined mainly to the magnetic field distribution, whereas, the  relativistic electrons are distributed more or less uniformly (Fig. \ref{cartoon}).
If this is indeed the case, it suggests $\delta' \approx 0$ in Eq. \ref{eq:cases}.

%

 \subsection{Combining with the multiwavelength Data}
 \label{multi}
 
 \begin{figure}
\centering
\includegraphics[width=0.68\textwidth]{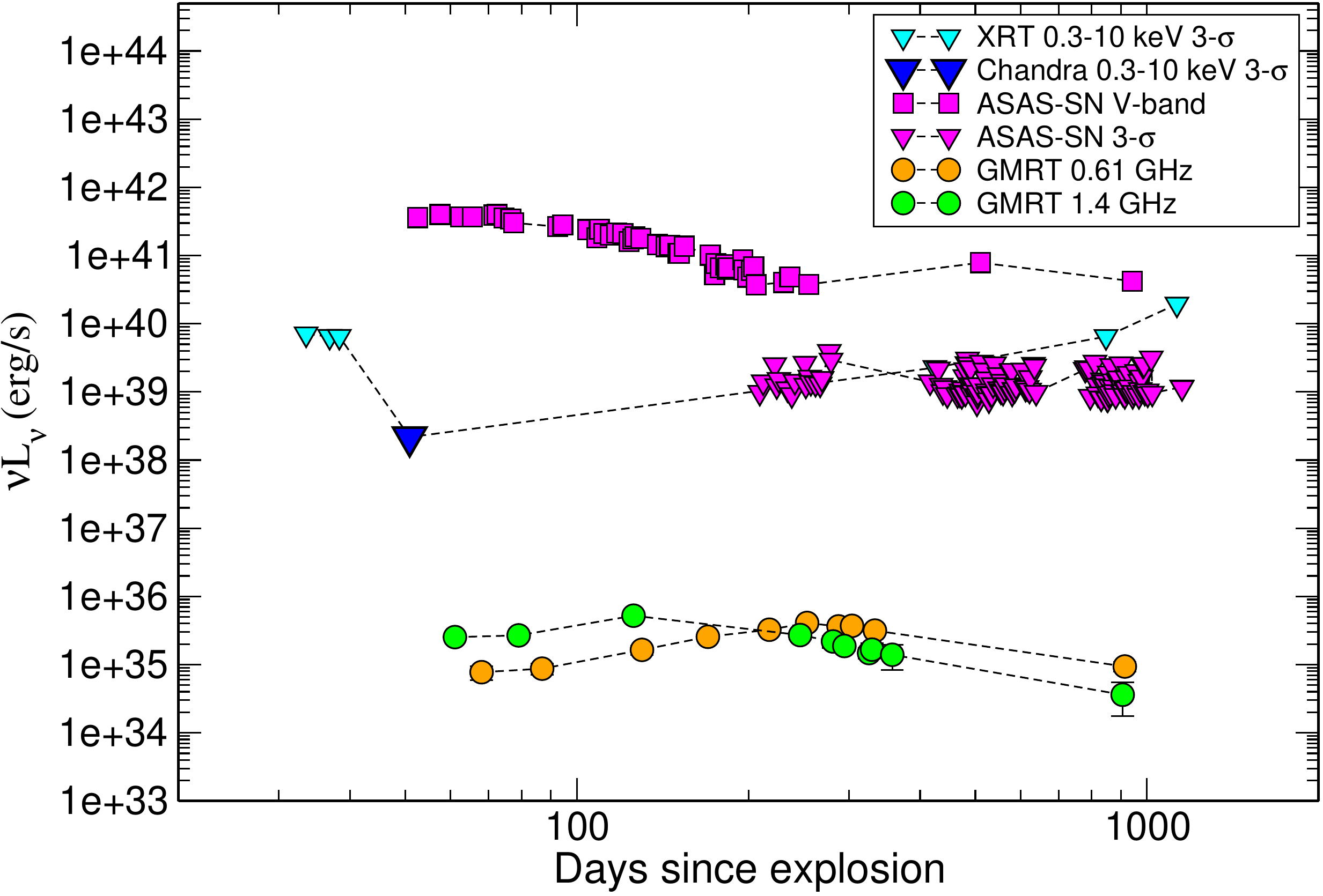}
\caption{A plot for $\nu L_\nu$ plot luminosities at various wavelengths against time. The SN is not detected in the X-ray bands.   {\it Chandra} observations around
day 51 (the bigger blue triangle) provide the most stringent upper limits on the X-ray luminosity.}
    \label{optical}
\end{figure}

The {\it Swift}-XRT data covered the epochs between days 33 and 1128 since explosion, however, the most important constraints come from the {\it Chandra} data on
day 51. 
At such early times, inverse Compton scattering of photospheric photons can contribute significantly to the X-ray emission in Type Ib/c SNe, and this prediction can be tested.

In Fig. \ref{optical}, we plot the radio, X-ray and optical luminosities for SN\,J1204. The X-ray luminosity is at least three orders of magnitude lower than the optical luminosity. 
 \citet{bjornsson13} estimated a  quantity $(L_{\rm x}/L_{\rm bol})^2/L_{\rm r}$, where $L_{\rm x}$, $L_{\rm bol}$ and $L_{\rm r}$ are the X-ray, bolometric and
 radio luminosities, respectively, 
normalized to  their respective values for another stripped envelope SN 2003L. They found  that this quantity  does not change significantly for various Type Ib/c SNe and remains close to 1, 
 even though the individual values of
 $L_{\rm x}$, $L_{\rm bol}$, and $L_{\rm r}$ are very different.  
 With the lack of X-ray detection as well as unavailability of measurement of bolometric luminosity in SN\,J1204, we cannot measure this parameter.
However, the observations with {\it Chandra} around day 51 gives the most constraining upper limits on the X-ray luminosity and the ASAS-SN data in the V-band can be treated as a lower limit on
 $L_{\rm bol}$. Using these values, we can constrain the above ratio, $(L_{\rm x}/L_{\rm bol})^2/L_{\rm r}$  for SN\,J1204, 
  scaled to the respective values for SN 2003L \citep{soderberg2006}, to be $(L_{\rm x}/L_{\rm bol})^2/L_{\rm r}<11.51$. This is indeed an upper limit and the actual value could be much smaller and  may fit in with the values obtained from other SNe Ib/c  \citep[Fig. 1 of][]{bjornsson13}.
 %
\citet{bjornsson13} also found this  quantity $(L_{\rm x}/L_{\rm bol})^2/L_{\rm r}$ to not evolve with time significantly,  
suggesting  that  $L_{\rm r} \propto  (L_{\rm x}/L_{\rm bol})^2$. This is consistent with a scenario in which a substantial contribution to the
 X-rays come from inverse Compton scattering of photospheric photons  \citep{bjornsson13}.
Thus inverse Compton scenario implies  an inhomogeneous source structure or vice-versa, if the equipartition fraction is not too far away from unity \citep[Eq. 22 of ][]{bjornsson13}.

\subsection{Final comprehensive  model}

 \begin{figure*}
\centering
\includegraphics[width=0.85\textwidth]{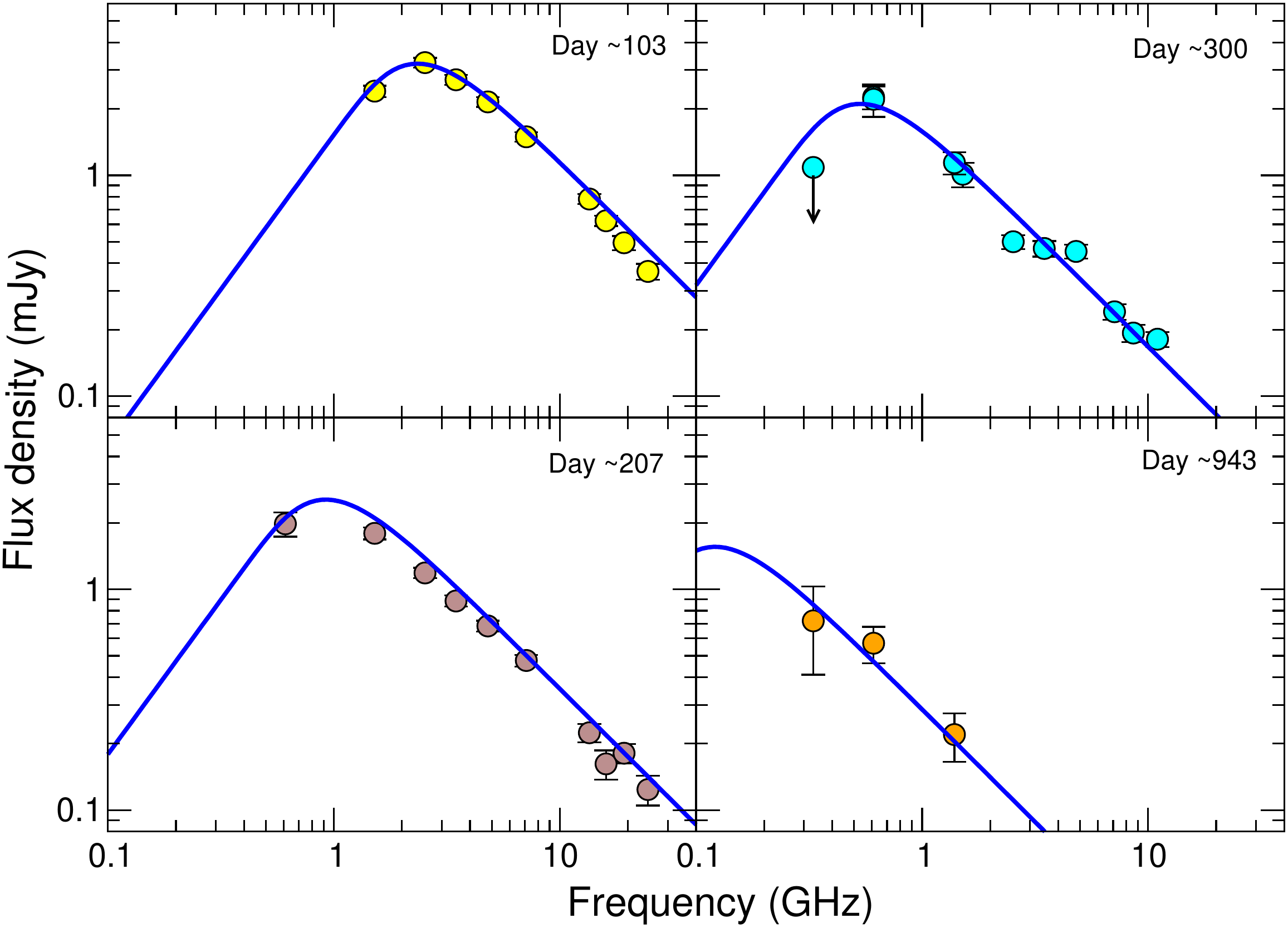}
\includegraphics[width=0.85\textwidth]{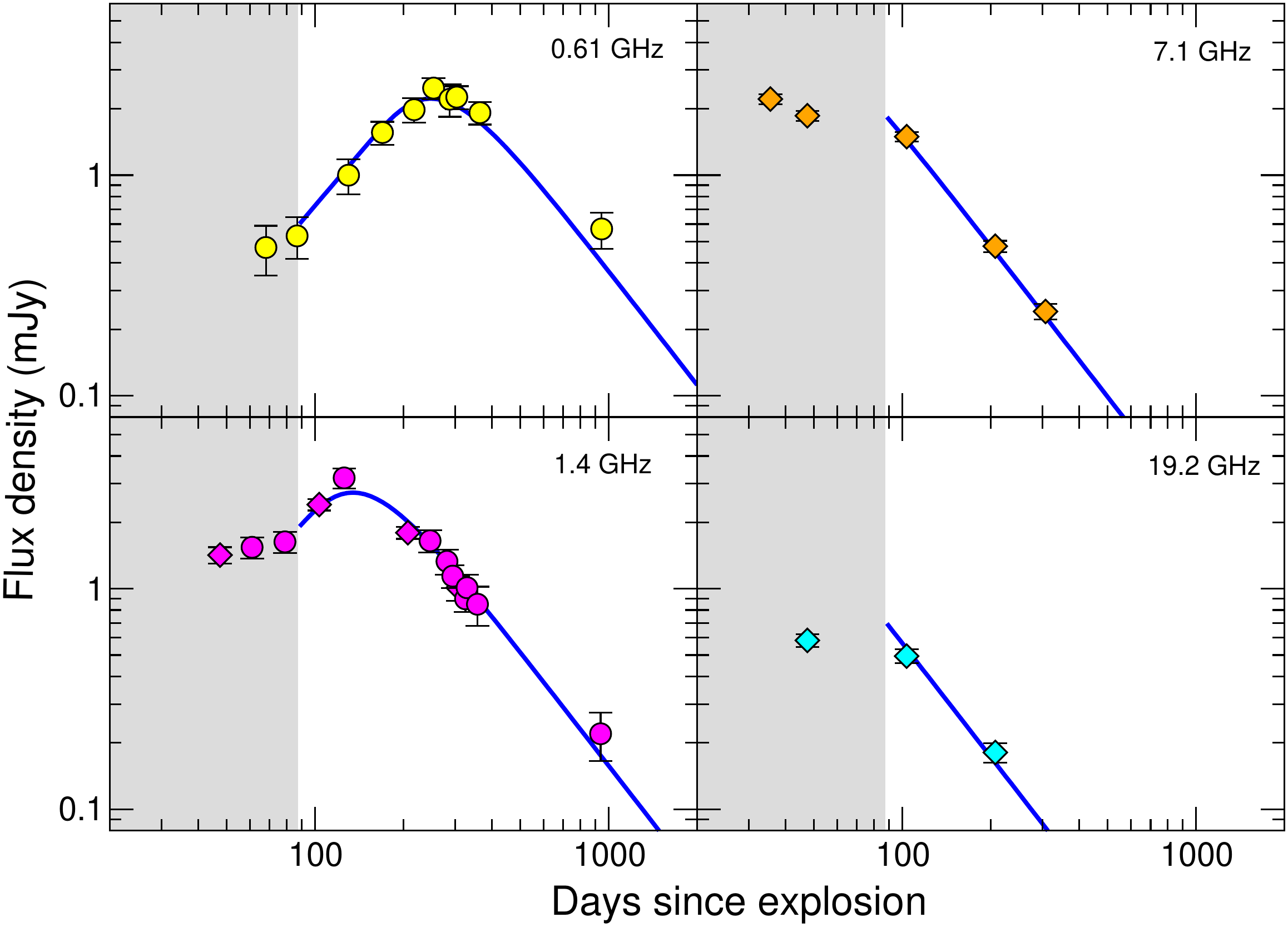}
\caption{The best fit inhomogeneous SSA model excluding data before 87 d. Here the shaded region in the light curves indicate the data excluded
from the modeling.  The best fit spectra (upper panel) and best fit light curves (lower panel) are plotted at four representative days and frequencies,
respectively.}
    \label{finalfit}
\end{figure*}

Connecting all the pieces together as discussed in the previous sections,  we establish that the radio emission from SN\,J1204  is arising from    the shock which has inhomogeneities, and is passing through a  shell during epochs $47$ 
to $87$ since explosion.
Hence we fit the inhomogeneous SSA model excluding all the
data till day 87.
%
We again use the formalism described in Eqns. \ref{eq:ssa1-inhomo} and \ref{eq:ssa2-inhomo}. 
The best fit model parameters are detailed in column 5 of Table \ref{fittedpara-gmrt}.  While the $\chi_\nu^2=1.6$ is still not a statistically very good fit, we consider it a reasonably acceptable fit.
We plot the spectra at four representative days and light curve at four representative frequencies in Fig. \ref{finalfit}. Here the shaded region in the light curves indicate the data excluded
from the modeling. These figures indicate that the inhomogeneous model fits the data reasonably well.  
%
  Unfortunately, it is difficult to quantitatively constrain the data before day 87.

\section{Discussion and Conclusions}
\label{discussion}

In this paper  
we have described the radio
observations of SN\,J1204 up to around 1200 days after the
explosion covering the frequency range from 0.33 GHz to 25 GHz.
The radio observations of SN\,J1204 suggest that the radio emitting region is inhomogeneous where the magnetic field is confined within a small distance from the shock front.
The data reveal that the  shock is passing through a  shell  during $\sim 47$ to $\sim 87$ days.

The GMRT low frequency data in the optically thick phase are crucial to indicate the presence of inhomogeneities in the synchrotron emitting region, which is responsible for the
flattening of the optically thick spectral index to $\sim 1.4$. 
Flattened light curves and spectra  have been seen in several SNe Ib/c, e.g., SNe 1994I \citep{weiler11}, 2003L \citep{soderberg2005} etc.  \citet{weiler11} explained the
flattened profile in SN 1994I due to  free-free absorption process  intrinsic to the synchrotron emitting source with the thermal electrons distributed roughly as the relativistic ones,
however, this scenario within the standard synchrotron model gives discrepant results \citep{bk17}.
While such an intrinsic FFA  mechanism is possible in Type IIn SNe due to their high densities \citep[e.g.][]{pc12}, it is very unlikely in SNe Ib/c which are expected to have Wolf-Rayet (W-R) progenitors.


 The ratio of $(L_{\rm x}/L_{\rm bol})^2/L_{\rm r}$ for SN\,J1204, where $L_{\rm x}$, $L_{\rm bol}$ and 
$L_{\rm r}$ are normalized to their respective values for SN 2003L , is $<11$.  However, 
it is an overestimation since we use  X-ray upper limit to  substitute for the
 X-ray luminosity and   ASAS-SN V-band luminosity as a proxy for bolometric luminosity.  
 Hence our value of  $(L_{\rm x}/L_{\rm bol})^2/L_{\rm r} < 11$ is not discrepant with other SNe Ib/c, indicating inverse-Compton
 effects likely to be important when the shock has inhomogeneities \citep{bjornsson13}.


 Since our  radio observations cover a large span of time, we have been able to fit the individual spectra up to around $\sim 1200$ days post explosion. 
The spectra between  days 47 and 103 suggest  rather insignificant time evolution in the shock parameters  (Fig. \ref{ssa-spectranblast wave para evolution}).  We have explained  this  in a scenario in which the shock is
 moving through a higher density shell between $\sim 47$ and $\sim 87$ days.
 We show in \S\ref{sec:shell} that the FFA optical depth in the shell is not expected to increase  by more than a factor of 2 and is likely to not alter the  radio  spectra significantly.
 Recently published optical spectroscopic data by \citet{singh+19} indicate no obvious new features associated with the shock impacting the shell. This indicates that the emission comes mainly from the ejecta and that the contribution from the shocked material is negligible. The lack of any signs of the shell-interaction would then be consistent with a rather low shell mass, i.e., a low FFA optical depth.

A major issue here is that when the shock is crossing the shell, one would assume the optically thin flux to increase in this duration due to continuous injection of electrons. This is contrary to what VLA data reveal, i.e. flatter optically thin light curves.  While this situation could have been  reconciled in the presence of  cooling, our data has suggested an absence of cooling.  
 We find that the early time flatter optically thin light curve evolution during the shell-crossing phase is consistent with a scenario where the  magnetic field distribution is confined
within a small region of the shock, whereas, the relativistic electrons are distributed more uniformly and the width of emission region stays nearly constant  (Fig. \ref{cartoon}).

SNe Ib/c are understood to be explosions of massive stars whose hydrogen envelopes have been stripped off before the explosion \citep{woosley2002}, 
but the physics of the process by which stars lose their outer envelopes, and the corresponding time scales are still debated \citep{smith2011}. 
 The presence of a shell during $\sim 47$ to $\sim  87$ days in SN\,J1204  
could be a clue to the stripping of the hydrogen/helium shell of the progenitor, possibly from a binary system 
\citep{podsiadlowski1992}.  Evidence of this has been seen in  for SNe Ib/c SN 2001em \citep{soderberg2004} and  SN 2014C \citep{vinko2017}, where evidence of Balmer recombination lines have been seen. In SNe 2001em and 2014C, the flux density enhancement was observed at a late time, i.e. $\sim$ day 677 for SN 2001em \citep{stockdale2005, chugai2006} and $\sim$ day 400 for SN 2014C \citep{anderson2017}. 
This suggested that the shells were ejected 
several decades before the explosion. 
 However, in case of SN\,J1204 we find the evidence of dense shell just 47 days after the explosion, lasting for $\le 40$ days. 
 Due to lack of constraints on ejecta and wind velocity,
we cannot determine the pre-explosion epoch at which shell was ejected, but it could not have been too long ago, unlike SNe 2001em and 2014C,  as W-R are known to have fast winds.
We have checked the archival ASAS-SN data to search for possible signatures of pre-SN ejection. 
 The ASAS-SN V-magnitude photometric observations begin $\sim$ 600 days pre-SN and do not show evidence of pre-SN ejection episodes (Fig. \ref{optical1}), however, there is no data at many epochs.
 Recent campaigns to observe SNe within days of explosion have revealed narrow emission lines of high-ionization species in the earliest spectra of luminous SNe II of all subclasses. These flash ionization features indicate the presence of a high-density medium close to the progenitor star \citep{flash18}.  In our case, the SN was detected  after the  optical maximum and hence no such data are available. 


The optically thin spectral index remains close to $-1$ at all epochs. 
This would suggest $p\approx3$, which is steeper than expected ($p\approx2$) from the standard diffusive shock acceleration theory.
 In absence of cooling, this situation can be  reconciled if the diffusive shock acceleration  is so efficient that the whole process becomes nonlinear \citep{chevalierfransson2006}. 
 The prediction of such a process is a flatter $p$ profile with time, albeit evolving very slowly \citep{chevalierfransson2006}. This can be tested at late epochs
 low frequency observations.

 In \S \ref{sec:a3}, we have derived the evolution of covering factor $f_{\rm B, cov}$. For SN\,J1204, $\beta'-\alpha'$ is positive, indicating that the time evolution for  $f_{\rm B, cov}$ is positive (Eq. \ref{eq:spec}). 
 This means the inhomogeneities should smooth out if followed long enough. We are continuing to observe SN\,J1204 at GMRT frequencies, especially at 325 MHz band, and these observations will test the above hypothesis, and reveal whether the synchrotron emitting region has emerged into a homogeneous one at late epochs.

To summarise the main conclusions of this work, the radio frequency observations of SN\,J1204 have revealed that  the radio emission is arising from a shock with
inhomogeneities mainly in the magnetic field distribution behind the shock (as sketched in Fig. \ref{cartoon}).
This shock is passing through a higher density shell for during $\sim47$ to $\sim  87 $ days. 
Low frequency sensitive telescopes like GMRT provide excellent opportunity to carry out such low frequency studies to reveal the 
nature of synchrotron emitting regions. With three times increased sensitivity as well as the near continuous and low-frequency wide bands of the upgraded GMRT \citep{gupta17}, such studies will be possible for a large number of SNe in the future.

\acknowledgments

We thank the referee for constructive comments, which helped improve the manuscript significantly. 
We acknowledge substantial help from Subo Dong, Jose L. Prieto, Krzusztof Z. Stanek, Christopher Kochanek, Todd Thompson and Tom Holoien for the ASAS-SN data,
and  Roger A. Chevalier.
P.C. acknowledges support from the Department of Science and Technology via SwaranaJayanti Fellowship award (file no.DST/SJF/PSA-01/2014-15). 
A.R. acknowledges Raja Ramana Fellowship of DAE, Govt of India. We thank the staff of the GMRT that made these observations possible. GMRT is run by the National Centre for Radio Astrophysics of the Tata Institute of Fundamental Research.
The National Radio Astronomy Observatory is a facility of the National Science Foundation operated under cooperative agreement by Associated Universities, Inc.
This research has made use of data obtained through the High Energy Astrophysics Science Archive Research Center Online Service, provided by the NASA/Goddard Space Flight Center.

\vspace{5mm}
\facilities{Giant Metrewave Radio Telescope, Karl J. Jansky Very Large Array, {\it Chandra} X-ray Telescope, {\it Swift} X-ray Telescope }

\appendix

\renewcommand\thefigure{\thesection.\arabic{figure}}
\setcounter{figure}{0}

\section{Inhomogeneous spherically symmetric model}
\label{sec:a1}

\begin{figure}
\centering
\includegraphics*[width=0.49\textwidth]{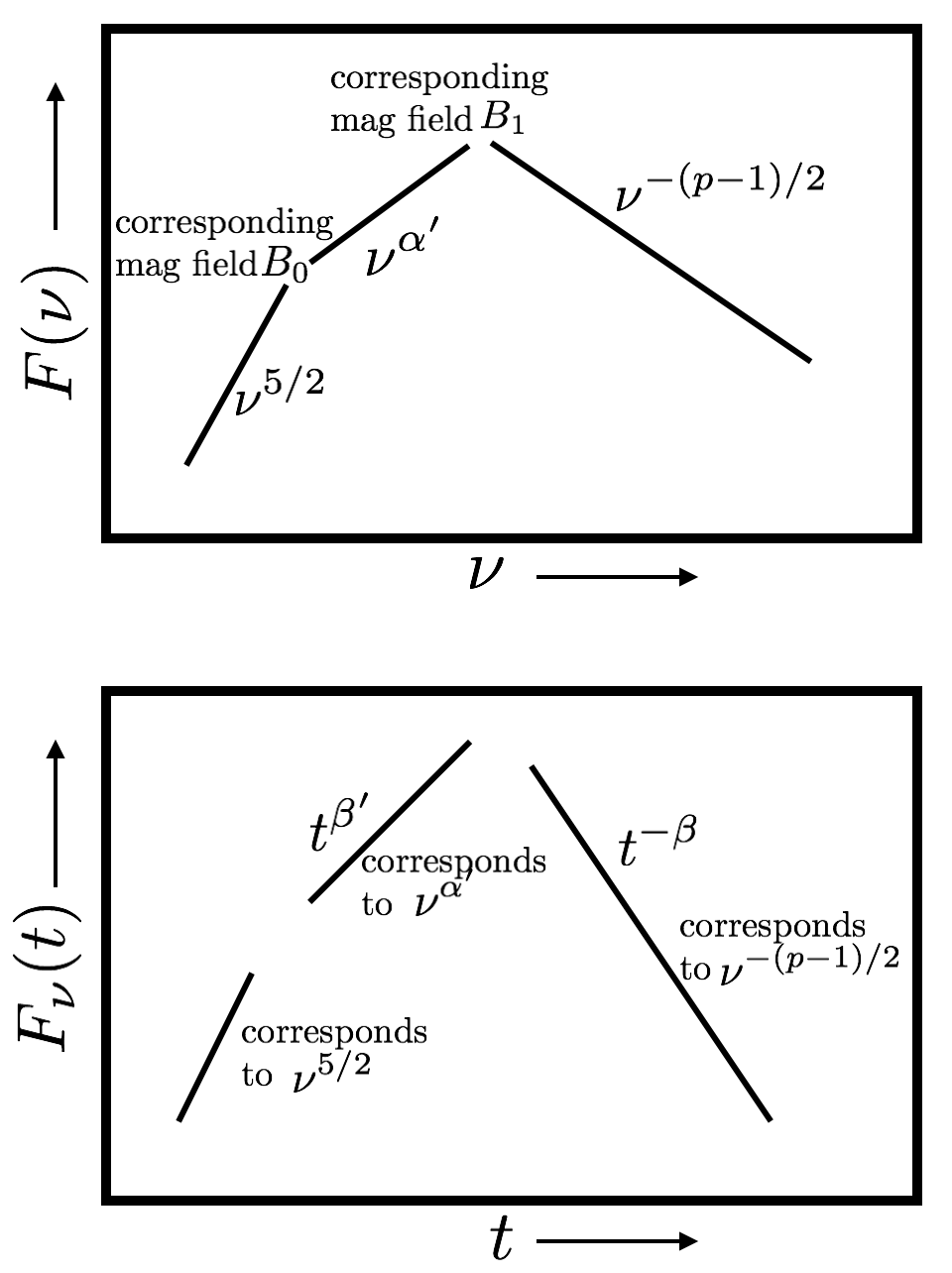}
\caption{ Cartoon diagram of a SN spectrum (above panel) for an inhomogeneous model. For $\nu<\nu_{\rm abs}(B_0)$, the spectrum will follow the
standard synchrotron model in optically thick phase ($F_\nu \propto \nu^{5/2}$). At the same time frequencies larger than the
frequency corresponding to SSA at $B_1$ (magnetic field at the peak), i.e. $\nu>\nu_{\rm abs}(B_1)$; follow standard synchrotron model in the optically thin phase
($F_\nu \propto \nu^{-(p-1)/2}$). Between these two boundaries, the spectrum will be affected by inhomogeneities and will follow  $F_\nu \propto \nu^{\alpha'}$ ($\alpha'<5/2$). The corresponding  light curve (lower panel)  will also gets modified accordingly and show flatter top.
}
    \label{fig:appendix}
\end{figure}       

In a homogeneous spherically symmetric model for Type Ib/c supernovae (SNe Ib/c), the radio emission is usually fit with a synchrotron emission model, suppressed at early times  by synchrotron self-absorption (SSA) with a frequency dependence of the flux density as  $F_\nu \propto \nu^{2.5}$. However,  inhomogeneities in an otherwise spherically symmetric emission structure can 
cause broadening in the observed radio spectra and/or light curves  \citep{bjornsson13}.
SSA frequency is quite sensitive to the presence of inhomogeneities and can be used to identify these in
the source structure.

The inhomogeneities can arise by variations in the  relativistic electrons distribution and/or the magnetic field strength within the synchrotron source.
\citet{bk17} have derived the formalism for the inhomogeneous synchrotron source, assuming planar geometry,  in terms of the covering factor $f_{B,\rm cov}$ 
characterizing the optically thick properties  and the filling factor $f_{B,\rm vol}$ characterizing the  optically thin properties of the radio emission. 
Here we discuss the relevant formalism taken mainly from \citet{bjornsson13, bk17}
in the context of this paper. 
Here it is assumed that the locally emitted spectrum is that of standard synchrotron model, however, the inhomogeneities in the
magnetic field ($B$) will give rise to variation in optical depths and superposition of spectra with varying optical depths will  broaden the resulting spectrum.

The source covering factor $f_{B,\rm cov}$ has been defined to describe the variation of the average magnetic field strength over the 
projected source surface. If $P(B)$ is the probability
of finding a particular value of $B$ within $B$ and $B+dB$, then the source covering factor will be parameterized from
$P(B) \propto B^{-a}$ and can be written as
\begin{equation}
f_{B, \rm cov } \approx f_{B_0, \rm cov } \left( \frac{B}{B_0} \right)^{1-a}
\label{eq:a1}
\end{equation}
Here we define magnetic field $B_0$ such that frequencies smaller than the frequency corresponding to SSA at $B_0$, i.e. $\nu<\nu_{\rm abs}(B_0)$
follow standard synchrotron model in optically thick phase, and $B_1$ such that   frequencies larger than the
frequency corresponding to SSA at $B_1$, i.e. $\nu>\nu_{\rm abs}(B_1)$ follow standard synchrotron model in the optically thin phase
 (Fig. \ref{fig:appendix}). Between these two boundaries, the covering factor will  modify the spectral flux as
  \begin{equation}
F(\nu) \propto \frac{R^2 \nu^{5/2} f_{B,\rm cov}}{B^{1/2}}
\end{equation}
The frequency dependence follows \citep{bk17}:
 \begin{equation}
\nu^3 \approx \nu^3_{\rm abs}(B) \propto U_eU_B r\left(\frac{\gamma_{\rm min}}{\gamma}\right)^{p-2}
\end{equation}
where $U_e$ and $U_B$ are energy densities of electrons and magnetic field, respectively, and $\gamma_{\rm min}$ is the minimum Lorentz factor such that
$N(\gamma){\rm d}\gamma \propto \gamma^{-p} {\rm d} \gamma$ for $\gamma\ge \gamma_{\rm min}$.
 It is also possible that the inhomogeneities in the relativistic electron distribution are correlated with the inhomogeneities in the magnetic field distribution. 
 \citet{bk17} defines a parameter $\delta'$ which indicates a possible correlation between the inhomogeneous distribution of 
 relativistic electrons  with the distribution of magnetic field strengths,  $U_e r \gamma_{\rm min}^{p-2} \propto B^{\delta'}$. 
  This gives
 \begin{equation}
 \nu_{\rm abs}(B) =\nu_{\rm abs}(B_0)  \left( \frac{B}{B_0}\right)^{\frac{p+2(1+\delta')}{p+4}} 
 \end{equation}
 \label{eq:aa4}
 Here $\delta'=0$ if no correlation exists between the two.
 
 Combining the above, between $\nu_{\rm abs}(B_0)< \nu < \nu_{\rm abs}(B_1)$, the spectral flux density can be written as 
   \begin{equation}
F(\nu) \approx  F(\nu_{\rm abs}(B_0)) \left( \frac{\nu}{\nu_{\rm abs}(B_0)}\right)^{\frac{3p+7+5\delta'-a(p+4))}{p+2(1+\delta')}} \equiv F(\nu_{\rm abs}(B_0)) \left( \frac{\nu}{\nu_{\rm abs}(B_0)}\right)^{\alpha'}
\label{eq:a5}
\end{equation}
 
 Thus in case of inhomogeneous emission structure, the spectral flux density can be defined as
 \begin{equation}
F(\nu)\propto \begin{cases}
\nu^\frac{5}{2},
     & \nu < \nu_{\rm abs}(B_0)    \\
   \nu^{\alpha'}\, \rm where \, \alpha'= {\frac{3p+7+5\delta'-a(p+4)}{p+2(1+\delta')}},
     &  \nu_{\rm abs}(B_0)< \nu < \nu_{\rm abs}(B_1)\\
     \nu^\frac{-(p-1)}{2},
      & \nu > \nu_{\rm abs}(B_1)  \: , 
\end{cases}
\label{eq:spec}
\end{equation}
The condition of $\alpha'\le 5/2$ and $\alpha \ge -(p-1)/2$ indicates
$1/2<a<(p+3)/2+\delta'$. Outside this range, the spectra is that of homogeneous source.

  For inhomogeneous model, the size of the radio emitting region $R$ cannot be determined simply from the observed peak and frequency $F(\nu_{\rm abs}(B_0))$, ${\nu_{\rm abs}(B_0)}$. 
 One needs to substitute the homogeneous model equivalent peak flux  $F_{\rm homo}(\nu_{\rm abs}(B_0))$ with 
 \begin{equation}
 F_{\rm homo}(\nu_{\rm abs}(B_0))=F(\nu_{\rm abs}(B_0))/f_{B_0,\rm cov}
 \label{eq:a7}
 \end{equation}
which gives
 \begin{equation}
R_p =8.8\times10^{15} \left( \frac{U_e}{U_B}\right)^{-\frac{1}{19}}  \left( \frac{F(\nu_{\rm abs}(B_0))}{\rm Jy} \right)^{\frac{9}{19}} 
\left( \frac{f_{B_0,\rm cov}}{1}\right)^{-\frac{9}{19}} \left( \frac{D}{\rm Mpc} \right)^{\frac{18}{19}}  \left( \frac{{\nu_{\rm abs}(B_0)}}{5\, \rm GHz} \right)^{-1}\rm cm
\label{eq:r}
\end{equation}
Covering factor $f_{B_0,\rm cov}$ poses large uncertainty, however, for spatially resolved inhomogeneous sources  covering factor can be derived from the brightness temperature $T_b$ \citep[$f_{B_0,\rm cov}(\nu) \propto F(\nu) \nu^{-2} T_b^{-1}$,][]{bk17},
where
$T_b \propto \nu^{\frac{\delta-1}{p+2(1+\delta)}}$
  in the transition region.

One can see that the  inhomogeneous model comes at the cost of several extra parameters, i.e. $f_{B_0,\rm cov}$, $a$, $B_1/B_0$, and $\delta'$, which is hard to constrain despite well-sampled radio observations, 
especially for unresolved sources for which such extra parameters are degenerate towards spectral width of the transition region.

\section{Evolution of SSA flux density and frequency}
\label{sec:a2}

From previous section (\S \ref{sec:a1}), flux density in an inhomogeneous model is
\begin{equation}
F(\nu) = F({\nu_{\rm abs} ( B_1))} \left( \frac{\nu}{\nu_{\rm abs}(B_1)}\right)^{\alpha'}
\end{equation}
One can determine the time evolution of the peak frequency and flux density  from it.
The above gives, for the optically thick part
\begin{equation}
F(\nu, t) = F(\nu_{\rm abs}(B_1), t) \times \nu_{\rm abs}(B_1, t)^{-\alpha'} \propto t^{\beta'}
\label{eq:b10}
\end{equation}
And likewise for the optically thin part
\begin{equation}
F(\nu,  t) = F(\nu_{\rm abs}(B_1), t) \times \nu_{\rm abs}(B_1, t)^{(p-1)/2} \propto t^{-\beta}
\end{equation}
Combining eqns \ref{eq:aa4} and \ref{eq:a5} give
\begin{eqnarray}
 \nu_{\rm abs}(B_1, t) \propto t^{-\frac{\beta'+\beta}{ \alpha'+\frac{p-1}{2}}} \\ \nonumber 
 F(\nu_{\rm abs}(B_1), t)  \propto t^{\beta'-\alpha'\frac{\beta'+\beta}{ \alpha'+\frac{p-1}{2}}}
 \label{eq:inhomo}
\end{eqnarray}

Since decline of $ \nu_{\rm abs}(B_1, t) $ is very fast, it is clear that $B_1$ should decline much faster than $t^{-1}$ or $R^{-1}$. Hence $B_1$ does not follow standard scalings, while
$B_0$ may still follow them.

\section{Evolution of the covering factor}
\label{sec:a3}

In order to determine the  time evolution of $f_{B_0,\rm cov}$, some assumption are needed.
Using Eq. \ref{eq:b10},

\begin{equation}
F({\nu_{\rm abs}(B_0)}, t)\propto t^{\beta'}  \nu_{\rm abs}(B_0,t)^{\alpha'} 
\label{eq:c14}
\end{equation}
For simplicity we assume $U_e \propto U_B$. Since $F({\nu_{\rm abs}(B_0)}, t)=f_{B_0,\rm cov}(t)  F_{\rm homo}({\nu_{\rm abs}(B_0)}, t)$ 
where $F_{\rm homo}({\nu_{\rm  abs}(B_0)})$ is the flux density corresponding to a homogeneous emitting region.
We can consider two cases; the magnetic and relativistic electron energy densities are proportional to the total post- shock energy density, i.e. $B_0 \propto t^{-1}$; and
another case of $B_0 \propto R^{-1}$ where the magnetic field corresponds to a case  when the  energy density is  inversely proportional to its  radiating surface

\begin{enumerate}
\item
Case A):
 $B_0 \propto t^{-1}$\\
From Eq. 5 of \citet{chevalier1998}: $\nu_{\rm abs}(B_0, t) \propto B_0 \propto t^{-1-2\frac{(1-m)}{(p+4)}} $ and $F_{\rm homo}({\nu_{\rm  abs}(B_0)}, t) \propto t^{-(1-m)\frac{(2p+13)}{(p+4)}}$
Hence using eqns. \ref{eq:a7} and \ref{eq:c14}\\
$f_{B_0,\rm cov} (t) \propto t^{\beta'} \nu_{\rm abs}(B_0,t)^{\alpha'} F_{\rm homo}({\nu_{\rm  abs}}(B_0), t)^{-1}$
gives
\begin{equation}
f_{B_0,\rm cov} (t) \propto t^{\beta'-\alpha'-2\alpha'\frac{(1-m)}{(p+4)}+(1-m)\frac{(2p+13)}{(p+4)}} \propto  t^{\beta'-\alpha'+(1-m)\frac{2(p-\alpha')+13}{p+4}}
\end{equation}

\item  Case B): 
$B_0 \propto R^{-1}$ ($R\propto t^m$),\\
From Eq. 6 of \citet{chevalier1998}: $\nu_{\rm abs}(B_0,t) \propto t^{-m} $ and $F_{\rm homo}({\nu_{\rm  abs}(B_0)}, t) \propto t^0$.
Hence using eqns. \ref{eq:a7}  and \ref{eq:c14}\\
\begin{equation}
f_{B_0,\rm cov} (t) \propto t^{\beta'} (t^{-m})^{\alpha'} \propto  t^{\beta'-\alpha'm} 
\end{equation}

\end{enumerate}

Thus
\begin{equation}
f_{B_0,\rm cov} (t) \propto \begin{cases}
t^{\beta'-\alpha'+(1-m)\frac{2(p-\alpha')+13}{p+4}},
     &  B_0 \propto t^{-1} \\
     t^{\beta'-\alpha'+(1-m)\alpha'},
      &  B_0 \propto R^{-1}  \propto  t^{-m} \: , 
\end{cases}
\label{eq:spec}
\end{equation}

If $\beta'>\alpha'$, the exponent is always positive, then one can see that $f_{B_0,\rm cov} (t) $ increases with time and hence inhomogeneities decrease with time irrespective of case A or B. Hence at late epochs an inhomogeneous 
model is expected to make a transition into a homogeneous model for such cases.

\clearpage

\end{document}